\documentclass[showpacs,preprintnumbers,amsmath,amssymb,aps,prd,nofootinbib,unicode-math]{revtex4-2}
\pdfoutput=1
\usepackage{graphicx}
\usepackage{amsmath,amssymb}
\usepackage{epstopdf}
\usepackage{preamble}

\graphicspath{{./Figures/}}

\usepackage{color}%

\usepackage[colorlinks,urlcolor=blue,citecolor=blue]{hyperref}
\def\hhref#1{\href{http://arxiv.org/abs/#1}{arXiv:#1}} 

\begin{document}

\title{Resurgence of the Effective Action in Inhomogeneous Fields}


\author{Gerald V. Dunne and Zachary Harris}
\affiliation{Physics Department, University  of Connecticut, Storrs, CT 06269}

\begin{abstract}

We show how background field inhomogeneities modify the non-perturbative structure of the effective action. The simple Borel poles of the Euler-Heisenberg effective action become branch points, and new branch points also appear, indicating new non-perturbative effects. This information is resurgently encoded in the perturbative weak field expansion, and becomes physically significant for strongly inhomogeneous fields.
We also show that resurgent extrapolation methods permit the decoding of a surprising amount of non-perturbative information from a relatively modest amount of perturbative input, enabling accurate analytic continuations from weak field to strong field, and of a spatially dependent magnetic background to a time dependent electric background. These extrapolations are far superior to standard WKB and locally constant field approximations.

\end{abstract}


\maketitle

\section{Introduction}
\label{sec:intro}

The one-loop QED effective action encodes the nonlinear and non-perturbative physics of the effective dynamics of the photon field, after integrating out the heavy electron and positron fields \cite{dittrich-reuter,Dittrich:2000zu}. It was first computed for electrons and positrons in the background of a constant electromagnetic field by Euler and Heisenberg \cite{euler} and Weisskopf \cite{weisskopf}, and later formalized for more general background fields by Schwinger \cite{schwinger,dunne-kogan}. One of its interesting applications is that the imaginary part of the effective action yields the rate of electron-positron pair production from the QED vacuum, due to the application of an external electric field (the ``Schwinger effect''). While this effect is too weak to have been directly observed, related QED processes approaching this ultra-high intensity regime are the subject of current experimental proposals \cite{Ringwald:2001ib,Yakimenko:2018kih,meuren,luxe,facet}, and correspondingly there has been a great deal of recent theoretical activity concerning strong inhomogeneities and also higher loops 
\cite{DiPiazza:2011tq,Fedotov:2016afw,Podszus:2018hnz,Ilderton:2019kqp,Karbstein:2019wmj,Mironov:2020gbi,Huet:2020awq,Gonoskov:2021hwf,Fedotov:2022ely,Heinzl:2022lly}. It is difficult to compute in this high intensity regime, very far from the natural regimes of perturbation theory or expansions around constant field approximations, and beyond the one-loop approximation (in the fine structure constant). Recently we applied methods of resurgent asymptotics and resurgent extrapolation to the QED effective action at two-loop order  for a constant background field, based on the pioneering work of Ritus \cite{ritus1,ritus2,ritus3}. We demonstrated that a remarkable amount of non-perturbative information can be efficiently decoded from a modest amount of perturbative information, opening a potential new approach to higher loop order computations \cite{Dunne:2021acr}. Here we apply similar ideas to the one-loop QED effective action in an {\it inhomogeneous} background field. We show that several new non-perturbative effects arise, and these play a crucial role for strongly inhomogeneous fields. These new effects are missed by the usual WKB and locally-constant-field approximations.

Since the perturbative weak field expansion of the effective action is asymptotic, it is naturally described by a Borel integral representation \cite{euler}, in which the non-perturbative physics is characterized by the singularities of the Borel transform. For the constant field Euler-Heisenberg case these Borel singularities are all simple poles, but we show here that field inhomogeneities change these poles into branch points, and also introduce new Borel singularities. These new features become especially significant as the field becomes more inhomogeneous. They have the physical consequence that the familiar exponentially small ``instanton'' effects in an electric background field acquire fluctuation series, so the effective action becomes a non-trivial trans-series.
These fluctuations are resurgently encoded in the coefficients of the perturbative expansion, as functions of the inhomogeneity scale. This is an example of the ``Cheshire Cat'' phenomenon, whereby resurgence relations may appear hidden in highly symmetric situations, but can be revealed by small perturbations \cite{Dunne:2016jsr,Kozcaz:2016wvy}. Finally, we show that resurgent extrapolation is able to produce accurate nonlinear and non-perturbative information based on a relatively small amount of perturbative data. In particular it is far superior to the commonly used WKB and locally-constant-field approximations, particularly for strongly inhomogeneous fields.

We first briefly recall the basics of the Euler-Heisenberg result, and the WKB and locally-constant-field (LCF) approximations. For a constant magnetic field $B$, the one-loop QED effective Lagrangian can be written in closed-form as a Borel integral \cite{euler,weisskopf,schwinger,dunne-kogan}:
\begin{equation}
\mathcal{L}_{\rm EH}\left(\frac{e B}{m^2}\right)=
- \frac{m^4}{8\pi}\qty(\frac{eB}{\pi m^2})^2\int_0^\infty \frac{\dd{s}}{s^2}\left(\coth (\pi s)-\frac{1}{\pi s}-\frac{\pi s}{3}\right)e^{-\pi m^2 s/(eB)} 
\label{eq:l1b}
\end{equation}
For a constant electric field, of magnitude $E$, the expression involves the real (``medianized" principal parts) integral, plus an exponentially suppressed imaginary part which is associated with the pair production rate:
\begin{equation}
\mathcal{L}_{\rm EH}\left(\frac{e E}{m^2}\right)=
 \frac{m^4}{8\pi}\qty(\frac{eE}{\pi m^2})^2 {\mathcal P}\int_0^\infty \frac{\dd{s}}{s^2}\left(\cot (\pi s)-\frac{1}{\pi s}+\frac{\pi s}{3}\right)e^{-\pi m^2 s/(e E)} + i\,  \frac{m^4}{8\pi}\qty(\frac{eE}{\pi m^2})^2\, \sum_{k=1}^\infty \frac{1}{k^2}\, e^{-k \pi m^2/(eE)}
\label{eq:l1e}
\end{equation}
where ${\mathcal P}$ denotes the Cauchy principal part integral.

The perturbative weak magnetic field expansion is asymptotic \cite{euler,schwinger,dittrich-reuter,graffi,chadha,dunne-kogan}:
\begin{equation}
\mathcal{L}_{\rm EH}\left(\frac{e B}{m^2}\right)
\sim  \frac{m^4}{4\pi^2}\sum_{n=0}^\infty (-1)^n \Gamma(2n+2)\zeta(2n+4)\left(\frac{eB}{\pi m^2}\right)^{2n+4}, \qquad eB\ll m^2
\label{eq:l1b-weak}
\end{equation}
where $\zeta(s)$ is the Riemann zeta function.
For an electric field the only difference perturbatively is that the coefficients do not alternate in sign, and correspondingly there is a non-perturbative imaginary part. This example is analogous to Dyson's physical instability argument concerning the expected divergence of the loop expansion \cite{dyson} (but here referring instead to the one-loop weak field expansion).

Beyond the constant field case, one must generally use approximations. For example: (i) the locally constant field approximation (LCFA), in which one replaces the constant field in the Euler-Heisenberg effective Lagrangian by its inhomogeneous form, and then integrates over the inhomogeneity directions to obtain the effective action. This is adaptable and easy to implement for localized fields, but as we shall see it is not accurate for very inhomogeneous fields; (ii) the LCFA can be extended to a systematic gradient expansion, but it is technically difficult to compute to high orders and the gradient expansion is also generically divergent, so the effective action becomes a multiple sum, each of which is divergent \cite{Dunne:1999uy}; (iii) the non-perturbative imaginary part can be efficiently analyzed using WKB methods when the inhomogeneity is one-dimensional \cite{Keldysh:1965ojf,brezin,Popov:1972vrp,Marinov:1977gq,Kim:2000un}; (iv) the worldline instanton approach, based on a saddle expansion of Feynman's worldline representation of the effective action \cite{Morette:1951zz,affleck,wli1,wli2,wli3}.

Our aim here is to propose another approach, similar to that in \cite{Dunne:2021acr}, in which we work with perturbative information, and then resum it using ideas from resurgent extrapolation \cite{Costin:2020hwg,Costin:2020pcj,Costin:2021bay}. As a first precision test of this idea for inhomogeneous fields, we apply it to two well-known examples of inhomogeneous fields for which the effective action is soluble \cite{narozhny}. We take linearly polarized fields, pointing in the $z$ direction, but with an amplitude that varies only in the $x$ direction for the spatially inhomogeneous magnetic field case, or in the $t$ direction for the time dependent electric field case:
\begin{subequations}
\begin{align}
  B(x)&=B\sech^2\qty(\frac{x}{\lambda})\label{eq:B}\\
  E(t)&=E\sech^2\qty(\frac{t}{\tau})\label{eq:E}
\end{align}
\end{subequations}
The effective action can be computed for these fields because the spectral problem for the corresponding Dirac operator reduces to a hypergeometric equation. In \cite{narozhny} the effective action is expressed as an integral over momenta. In \cite{Cangemi:1995ee,Dunne:1997kw} these integrals were done, reducing the expression for the effective action to a single Borel-type integral: see Eq. \eqref{eq:exact2}.

We begin with the magnetic field case, and ask three main questions:
\begin{enumerate}
\item How does the resurgent trans-series structure of the effective action change due to the field inhomogeneity?
    \item Given some finite-order information about the perturbative weak magnetic field expansion, can we extrapolate efficiently to the strong magnetic field regime, even for very inhomogeneous fields?
    \item Given some finite-order information about the perturbative weak magnetic field expansion, can we analytically continue efficiently to the electric field regime, even for very inhomogeneous fields?
\end{enumerate}

\begin{figure}[htb]
\centerline{\includegraphics[scale=1]{"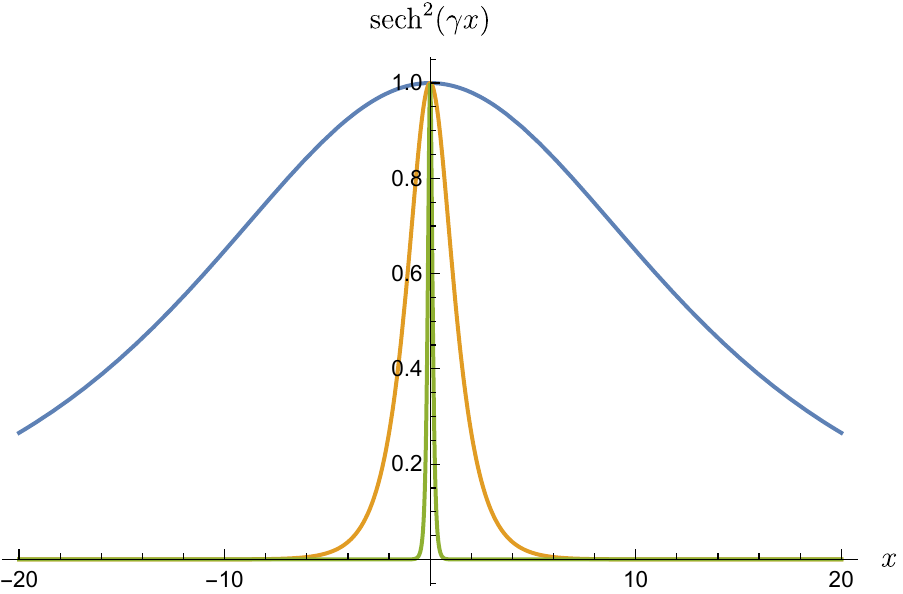"}}
\caption{Profiles of the inhomogeneous fields in \eqref{eq:B}-\eqref{eq:E}, for three different values of the inhomogeneity parameter: $\gamma=0.1$ (blue curve), $\gamma=1$ (orange curve), and $\gamma=10$ (green curve). Larger $\gamma$ corresponds to more strongly inhomogeneous field profiles.}
\label{fig:sech2}
\end{figure}
 The classical background fields in \eqref{eq:B}-\eqref{eq:E} are characterized by two parameters: an amplitude parameter, $B$ or $E$, and a scale parameter, $\lambda$ or $\tau$. The natural dimensionless parameter to describe the degree of field inhomogeneity is the ``Keldysh inhomogeneity parameter" \cite{Keldysh:1965ojf}:
\begin{equation}
  \gamma=\frac{m}{eB\lambda}\qquad \text{or} \qquad \gamma=\frac{m}{eE\tau}
  \label{eq:gamma}
\end{equation}
which measures the inhomogeneity scale ($\lambda$ or $\tau$) in terms of the classical scale set by the interaction of a magnetic or electric field with particles of charge-to-mass ratio $e/m$. See Figure \ref{fig:sech2}.
The inhomogeneity parameter $\gamma$ is analogous to a 't Hooft coupling: we compute in the effective field theory limit in which the electron mass sets the large scale: $m^2\gg B$ and $m\gg 1/\lambda$, while the product $B \lambda$ remains finite. However, $\gamma$ could be small (the homogeneous limit) or large (the inhomogeneous limit). 

The effective actions for these two configurations are related by the simultaneous analytic continuations
\begin{equation}
  B^2\mapsto -E^2\quad\text{and}\quad \lambda^2\mapsto-\tau^2\label{eq:AC}
\end{equation}
which can be implemented in such a way that $\gamma$ remains unchanged. Our strategy will be to start with the perturbative weak magnetic field expansion, for which the effective action has no imaginary part, and extrapolate and analytically continue from there, first from weak to strong magnetic field and then from weak magnetic field to weak and strong electric field. In order to do this we need as accurate as possible a Borel representation of the divergent weak $B$ field expansion, including its dependence on $\gamma$.

The general form of the perturbative weak $B$ field expansion for a linearly polarized magnetic field characterized by a single inhomogeneity parameter $\gamma$ is:
\begin{equation}
S(B,\lambda)\sim \lambda L^2T\,\frac{m^4}{3\pi^2}\sum_{n=0}^\infty a_n(\gamma)\qty(\frac{eB}{\pi m^2})^{2n+4} \quad ,\qquad eB\ll m^2
\label{eq:weak-field-exp}
\end{equation}
Here $L$ and $T$ denote the spatial and temporal extent of the {\it homogeneous} directions. 
The coefficients $a_n(\gamma)$ in \eqref{eq:weak-field-exp} are no longer pure numbers (as for the homogeneous field case in (\ref{eq:l1b-weak})), but depend on the inhomogeneity parameter $\gamma$. In general one could compute terms of this expansion, for chosen values of $\gamma$, and study the nature of this divergent weak field expansion. However, we can test this procedure more precisely for the fields in (\ref{eq:B})-(\ref{eq:E}) because there  exists a closed-form Borel-like integral representation of the effective action \cite{Cangemi:1995ee,Dunne:1997kw}:
\begin{eqnarray}
\begin{aligned}
  S(B,\lambda)&=\lambda L^2T\,\frac{m^4}{3\pi^2}\int_0^\infty \frac{\dd{s}}{e^{\pi m^2s/(eB)}-1}\Bigg[\dv{z}{s}\qty(1-\frac{4z}{3}+\frac{z^2}{5}\, _{2}F_{1}\qty(1,1;\tfrac{7}{2};z))-(s\to -s)\\
  &
  \hskip 5.5cm
  -\Bigg(\frac{8}{3}s-2\gamma \qty(1+\frac{\gamma^2}{4}s^2)^{3/2}\,\text{arcsinh}\qty(\frac{\gamma}{2} s)\Bigg)\Bigg]\label{eq:exact2}
\end{aligned}
\end{eqnarray}
where we have defined the convenient variable $z$ as
\begin{equation}
  z=-is-\frac{\gamma^2}{4} s^2
  \label{eq:z}
\end{equation}
The final term in (\ref{eq:exact2}) involves the zero field subtraction and the homogeneous field subtraction, enabling a smooth comparison at large and small $\gamma$. In previous work, this explicit integral representation has led to detailed analysis of the Borel summation of the gradient expansion \cite{Dunne:1999uy} and also the analytic continuation between the magnetic and electric background field configurations \cite{Dunne:1998ni}. Here we generate the weak field expansion to find explicit expressions
for the expansion coefficients $a_n(\gamma)$:
\begin{subequations}
\begin{align}
  a_0(\gamma)&=\frac{4\pi^4}{525}+\frac{2\pi^4}{75}\gamma^2
  \label{eq:an-gamma0}\\
  a_{n>0}(\gamma)&=\frac{3\sqrt{\pi}}{4}(-1)^n\Gamma(2n+4)\zeta(2n+4)\Bigg[\frac{\Gamma(2n+2)}{\Gamma\qty(2n+\tfrac{9}{2})}\, _{3}F_{2}\qty({\scriptstyle -2-n,-\tfrac{3}{2}-n,-\tfrac{7}{2}-2n \atop \scriptstyle -1-2n,-3-2n}\bigg|-\gamma^2)-\frac{2\Gamma(n)}{\Gamma\qty(n+\tfrac{5}{2})}\qty(\frac{\gamma}{2})^{2n+4}\Bigg]
  \label{eq:an-gamma}
\end{align}
\end{subequations}
Note that the $_{3}F_2$ functions truncate for integer $n$, so the coefficients $a_n(\gamma)$ are in fact polynomials of degree $(n+1)$ in $\gamma^2$.

\section{Large Order Behavior and Resurgence}
\label{sec:large-order}

In this Section we show how non-perturbative features of the effective action are encoded in the large-order growth behavior of the perturbative coefficients $a_n(\gamma)$ in \eqref{eq:an-gamma0}-\eqref{eq:an-gamma}. This is one of the indicators of resurgence structure, and here we demonstrate that the way it manifests itself changes quite dramatically when we generalize the Euler-Heisenberg result (which is for a homogeneous background field) to an inhomogeneous background field. We begin by recalling how this resurgence structure appears in various approximations, and then we turn to the full picture.

\subsection{Large-Order Behavior and Resurgence in the Euler-Heisenberg Constant Field Case}
\label{sec:large-he}

For the Euler-Heisenberg constant background field case \eqref{eq:l1b-weak}, the perturbative coefficients are pure numbers, with large-order growth being factorial, with an infinite series of exponential corrections:
\begin{eqnarray}
a_n^{\rm EH}
= (-1)^n \Gamma(2n+2)\, \zeta(2n+4)
=(-1)^n \Gamma(2n+2) \sum_{k=1}^\infty \frac{1}{k^{2n+4}}
\label{eq:he-large}
\end{eqnarray}
To understand the physical significance of the sum over exponential corrections $\frac{1}{k^{2n+4}}$, we can write the Borel transform function in (\ref{eq:l1b}) as a partial fraction expansion
\begin{eqnarray}
-\frac{1}{s^2}\left(\coth(\pi s)-\frac{1}{\pi s}-\frac{\pi s}{3}\right) = \frac{2}{\pi}\sum_{k=1}^\infty \frac{1}{k^2}\frac{s}{(s^2+k^2)}
\label{eq:he-partial}
\end{eqnarray}
Therefore we can write the Euler-Heisenberg effective Lagrangian \eqref{eq:l1b} directly as a sum
\begin{eqnarray}
{\mathcal L}_{\rm EH}\left(\frac{eB}{m^2}\right) =\frac{1}{4\pi^2} \sum_{k=1}^\infty \left(\frac{e B}{k\pi m^2}\right)^2 \int_0^\infty \dd{s}\, e^{-k\pi m^2 s/(eB)}\, \frac{s}{s^2+1}
\label{eq:he-ksum}
\end{eqnarray}
This form emphasizes the fact that the singularities of the Borel transform function (\ref{eq:he-partial}) are essentially identical simple poles, equally spaced along the imaginary Borel axis, and can be captured as a sum over just two symmetric poles at $s=\pm i$, as in (\ref{eq:he-ksum}). The expansion variable is the inverse of the square of the $k^{\rm th}$ ``instanton action'': ${\mathcal A}_k=\frac{k\pi m^2}{eB}$, for $k=1, 2, \dots$. As is well known, after rotating $B^2\to - E^2$, this yields the infinite sum over multi-instantons for the imaginary part of the effective Lagrangian in (\ref{eq:l1e}). This explains why there is only one large-order/low-order resurgence relation for the constant field Euler-Heisenberg expression, relating the large order growth (\ref{eq:he-large}) of the perturbative coefficients to the entire multi-instanton sum. The instanton terms have no fluctuation factors, but nevertheless each instanton term is indeed encoded in the perturbative series, including its scaled action ${\mathcal A}_k=\frac{k\pi m^2}{eE}$ and its prefactor $\frac{1}{{\mathcal A}_k^2}$. The reason for recalling this here is that a similar multi-instanton structure appears even for inhomogeneous background fields, but with some interesting additional new features.

\subsection{Large-Order Behavior and Resurgence in the Locally Constant Field Approximations}
\label{sec:large-lcfa}

A simple approximation method for inhomogeneous background fields is the locally constant field approximation (LCFA), which is just the integrated form of the leading order of the gradient expansion;  namely, the Euler-Heisenberg result (\ref{eq:l1b}) with the constant field replaced by its inhomogeneous form, then integrated over spacetime. For example, for the inhomogeneous magnetic and electric fields (\ref{eq:B})-(\ref{eq:E}) we obtain:
\begin{align}
  S_\text{LCFA}(B,\lambda)&=\lambda L^2T\int_{-\infty}^\infty \dd{x} \mathcal{L}_\text{EH}\qty(B\sech^2x)
  \label{eq:lcfa-magnetic}
   \\
   S_\text{LCFA}(E,\tau)&= L^3\tau\int_{-\infty}^\infty \dd{t} \mathcal{L}_\text{EH}\qty(E\sech^2 t)
   \label{eq:lcfa-electric}
\end{align}
In the magnetic case the corresponding weak field expansion is:
\begin{equation}
    S_\text{LCFA}(B,\lambda)\sim \lambda L^2 T\,\frac{m^4}{4\pi^{3/2}}\sum_{n=0}^\infty (-1)^n \Gamma(2n+2)\zeta(2n+4) \frac{\Gamma(2n+4)}{\Gamma(2n+\tfrac{9}{2})}\qty(\frac{eB}{\pi m^2})^{2n+4},\qquad eB\ll m^2\label{LCFAweakfield}
\end{equation}
This weak $B$ field expansion looks very similar to the Euler-Heisenberg expansion in (\ref{eq:l1b-weak}), but with coefficients being multiplied by a ratio of Gamma factors coming from the $x$ integration.\footnote{We have used the integral: $\int_{-\infty}^\infty \dd{x} \qty(\sech^2 x)^{2n+4}= \sqrt{\pi} \Gamma(2n+4)/\Gamma(2n+9/2)$.} The leading large-order growth of these coefficients as $n\to\infty$ is therefore given by:
\begin{eqnarray}
a_n^{\text{LCFA}}&=&(-1)^n \Gamma(2n+2)\zeta(2n+4) \frac{\Gamma(2n+4)}{\Gamma\qty(2n+\tfrac{9}{2})}
\nonumber
\\
&\sim & (-1)^n\Gamma\qty(2n+\tfrac{3}{2})\zeta(2n+4)\qty[1-\frac{5}{4}\frac{1}{\qty(2n+\tfrac{1}{2})}+\frac{105}{32}\frac{1}{\qty(2n+\tfrac{1}{2})\qty(2n-\tfrac{1}{2})}+\ldots]
\label{eq:lcfa-b}
\end{eqnarray}
This expression deserves several comments: 
\begin{itemize}
    \item 
    The {\it leading} factorial growth rate  has changed from $\Gamma(2n+2)$ in (\ref{eq:he-large}) to $\Gamma\left(2n+\frac{3}{2}\right)$ in \eqref{eq:lcfa-b}.
    \item
    The overall zeta factor $\zeta(2n+4)$ in \eqref{eq:lcfa-b}, which effectively encodes the multi-instanton sum, is the same as in \eqref{eq:he-large}. 
    \item
    The factorial growth in \eqref{eq:lcfa-b} is now multiplied by an infinite series of sub-leading power-law corrections. There are no such power-law corrections in the Euler-Heisenberg case \eqref{eq:he-large}.
    \item
    The coefficients of the subleading corrections in \eqref{eq:lcfa-b} are rational and grow factorially in magnitude:
\begin{eqnarray}
\left\{1, -\frac{5}{4}, \frac{105}{32}, -\frac{1575}{128},\frac{121275}{2048},-\frac{2837835}{8192}, \dots \right\} = 
\frac{4}{3\pi} \frac{(-1)^n\Gamma\left(n+\frac{1}{2}\right) \Gamma\left(n+\frac{5}{2}\right)}{n!} 
\label{eq:lcfa-flucs}
\end{eqnarray}
As an example of a large-order/low-order resurgence relation, these coefficients appear in the fluctuations about the instantons in equation \eqref{eq:lcfa} below.
\end{itemize}

For the inhomogeneous electric field in (\ref{eq:E}) the perturbative expansion is the same as  for the inhomogeneous magnetic field in (\ref{eq:lcfa-b}), but the coefficients do not alternate in sign. The imaginary part of the effective action in the LCFA can be computed from the imaginary part of the Euler-Heisenberg result (\ref{eq:l1e}), yielding a simple expression in terms of a confluent hypergeometric function $U(a,b;x)$: 
\begin{align}
\left[  \Im S(E,\tau)\right]_\text{LCFA} 
&= L^3\tau\frac{m^4}{8\pi}\sum_{k=1}^\infty \qty(\frac{eE}{k\pi m^2})^{\!2}\int_{-\infty}^\infty \dd{t} \sech^4(t)\, \exp\left(-\frac{k\pi m^2 }{eE} \cosh^2 t\right) 
\nonumber\\
& =  L^3\tau\frac{m^4}{8\pi^{1/2} }\sum_{k=1}^\infty \qty(\frac{eE}{k\pi m^2})^{\!2} \, U\left(\frac{1}{2}, -1; \frac{k\pi m^2 }{eE}\right)e^{-k\pi m^2/(eE)}
\nonumber\\
 &\sim 
   L^3\tau\frac{m^4}{8\pi^{1/2}}
   \sum_{k=1}^\infty
   \qty(\frac{eE}{k\pi m^2})^{5/2}e^{-k \pi m^2/(eE)}\qty[1-\frac{5}{4}\qty(\frac{eE}{k\pi m^2})+\frac{105}{32}\qty(\frac{eE}{k\pi m^2})^2+\ldots]
\label{eq:lcfa}
\end{align}
The final expression in \eqref{eq:lcfa} should be compared with the imaginary part of the Euler-Heisenberg result \eqref{eq:l1e}.
The LCFA changes quite dramatically the structure of the imaginary part of the effective action. In the constant field Euler-Heisenberg result (\ref{eq:l1e}) the instanton sum (the sum over $k$) is a sum of pure exponential factors, with no fluctuation terms. However, in the LCFA we see that each instanton term is multiplied by an asymptotic series of fluctuations. Furthermore, note that the fluctuation series for each instanton sector (labeled by the integer $k$) is identical when expressed in terms of the $k^{\rm th}$ instanton action $\frac{k\pi m^2}{eE}$. In addition, we recognize the expansion coefficients of the fluctuation series in (\ref{eq:lcfa}) from the subleading corrections of the large-order growth of the perturbative series coefficients for the magnetic field case in \eqref{eq:lcfa-b}. This is a typical large-order/low-order resurgence relation \cite{berry-howls,Ecalle,costin-book,Marino:2012zq,Dorigoni:2014hea,gokce} relating the fluctuations about different saddle point sectors: the perturbative expansion coefficients encode all the information about the fluctuations about each of the infinite tower of multi-instanton saddles. This manifestation of resurgence is absent in the constant field limit, because the associated fluctuation series truncate, but it is present when the inhomogeneity parameter becomes nonzero, even in the locally constant field approximation.

Another interesting way to probe this structure is to compute the Borel representation in the LCFA. For the magnetic case, applying the LCFA directly to the Borel integral \eqref{eq:l1b} yields:
\begin{equation}
S_{\rm LCFA}(B, \lambda)= 
-\lambda L^2T\,\frac{m^4}{8\pi^{1/2}}\qty(\frac{eB}{\pi m^2})^2\int_0^\infty \frac{\dd{s}}{s^2}\left(\coth(\pi s)-\frac{1}{\pi s}-\frac{\pi s}{3}\right) U\!\left(\frac{1}{2}, -1; \frac{\pi m^2 s}{eB}\right)e^{-\pi m^2s/(eB)}
\label{eq:borel-lcfa}
\end{equation}
This expression \eqref{eq:borel-lcfa} reproduces the LCFA weak magnetic field expansion \eqref{LCFAweakfield}, and the analytic continuation from magnetic to electric generates the full non-perturbative imaginary part in the LCFA \eqref{eq:lcfa}. 
The hypergeometric factor has no singularities in the finite plane,\footnote{Note that this hypergeometric U function reduces to a combination of modified Bessel K functions: $\frac{3\sqrt{\pi}}{2} e^{-s} U\left(\frac{1}{2}, -1; s\right)=e^{-s/2}\left(s^2 K_0\left(\frac{s}{2}\right)-s(s-1)K_1\left(\frac{s}{2}\right)\right)$, which has infinite radius of convergence.} so we see that the Borel singularities in the LCFA are exactly the same as for the Euler-Heisenberg case (recall the Borel representation \eqref{eq:l1b}). However, now the residues at these poles are not simply exponentials, as in the Euler-Heisenberg case \eqref{eq:l1b}, but give Bessel function terms, whose weak-field expansions produce exponentials multiplied by an asymptotic expansion, as in \eqref{eq:lcfa}. 

\begin{figure}[htb]
    \centering
    \includegraphics[width=\textwidth]{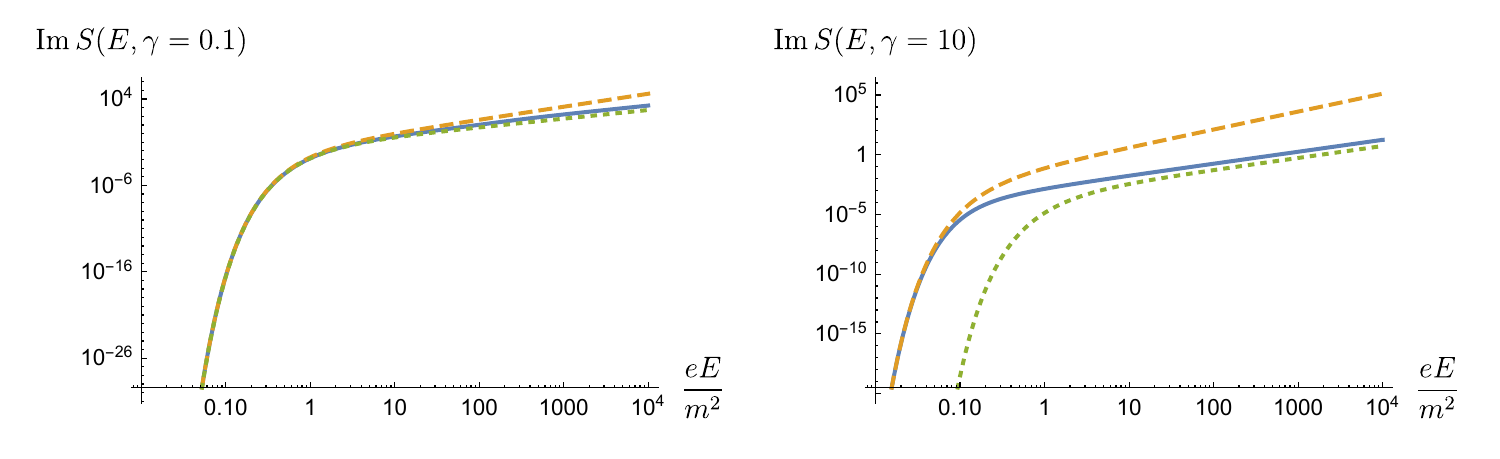}
    \caption{The exact imaginary part of the effective action in (\ref{eq:exact2}) after analytically continuing to the electric field background (\ref{eq:E}) (blue, solid) compared with the locally constant field approximation (\ref{eq:lcfa}) (green, dotted) and the standard WKB approximation (\ref{eq:wkb}) (orange, dashed), for both small $\gamma$ (left) and large $\gamma$ (right). The LCFA is only accurate at small $\gamma$, while the WKB approximation is only accurate at weak fields. These plots should also be compared with our resurgent extrapolations in Figure \ref{fig:PCB-electric}.}
    \label{fig:ImSapprox}
\end{figure}

In Figure \ref{fig:ImSapprox} we plot the leading ($k=1$) contribution to the LCFA expression \eqref{eq:lcfa} for the imaginary part of the effective action (green dotted curves), compared to the exact result \eqref{eq:exact2} (blue solid curves), for a small inhomogeneity parameter ($\gamma=0.1$, left) and for a large inhomogeneity parameter ($\gamma=10$, right). As expected, the agreement of the LCFA approximation with the exact result is better for smaller $\gamma$. Figure \ref{fig:ImSapprox} also shows the corresponding WKB approximations (orange dashed curves), as discussed in the next subsection.

\subsection{Large-Order Behavior and Resurgence in the WKB Approximation}
\label{sec:large-wkb}

The WKB approximation can be used to compute the leading imaginary part of the effective action in the presence of the inhomogeneous electric field (\ref{eq:E}) \cite{brezin,Popov:1972vrp,Marinov:1977gq}. This generalizes to QED the original treatment of atomic ionization by Keldysh \cite{Keldysh:1965ojf}. For this particular field one finds \cite{Marinov:1977gq,dunne-kogan}:
\begin{equation}
 \left[ \Im S(E,\tau)\right]_\text{WKB} \sim L^3 \tau\frac{m^4}{8\pi^{1/2}}\qty(\frac{eE}{\pi m^2})^{5/2}\qty(1+\gamma^2)^{5/4}\exp\qty(-\frac{\pi m^2}{eE}\frac{2}{\sqrt{1+\gamma^2}+1})
 \label{eq:wkb}
\end{equation}
Several comments are in order concerning the comparison with the constant field case and with the LCFA:
\begin{itemize}
\item
The exponent now has a $\gamma$ dependent factor, $\frac{2}{\sqrt{1+\gamma^2}+1}$, unlike either the Euler-Heisenberg result in \eqref{eq:l1e} or the LCFA result in \eqref{eq:lcfa}.
\item
The prefactor power $E^{5/2}$ of the electric field differs from Euler-Heisenberg, but is the same as in the LCFA \eqref{eq:lcfa}.
\item
The WKB expression \eqref{eq:wkb}  has a $\gamma$ dependent prefactor, $\qty(1+\gamma^2)^{5/4}$, which is of course absent for the constant field case, and is also absent for the LCFA.
\item
By construction, the standard WKB approximation \eqref{eq:wkb} only yields the {\it leading} exponential term, and does not sum over all multi-instantons, whereas both the Euler-Heisenberg \eqref{eq:l1e} and LCFA \eqref{eq:lcfa} expressions do sum over all multi-instantons.\footnote{This could be incorporated in the worldline instanton formalism, by summing over multiple windings of the periodic orbits \cite{Morette:1951zz,affleck,wli1,wli2}, but is not part of the conventional WKB approximation.}

\end{itemize}

In Figure \ref{fig:ImSapprox} we plot the WKB approximation \eqref{eq:wkb} (orange dashed curves), compared to the  exact result \eqref{eq:exact2} (blue solid curves) and the leading LCFA result (green dotted curves), for a small inhomogeneity parameter ($\gamma=0.1$, left) and for a large inhomogeneity parameter ($\gamma=10$, right). The WKB approximation is only good at weak field, with the discrepancy becoming more significant at larger $\gamma$; i.e. for fields that are more inhomogeneous.

\begin{figure}[b]
    \centering
    \includegraphics[width=\textwidth]{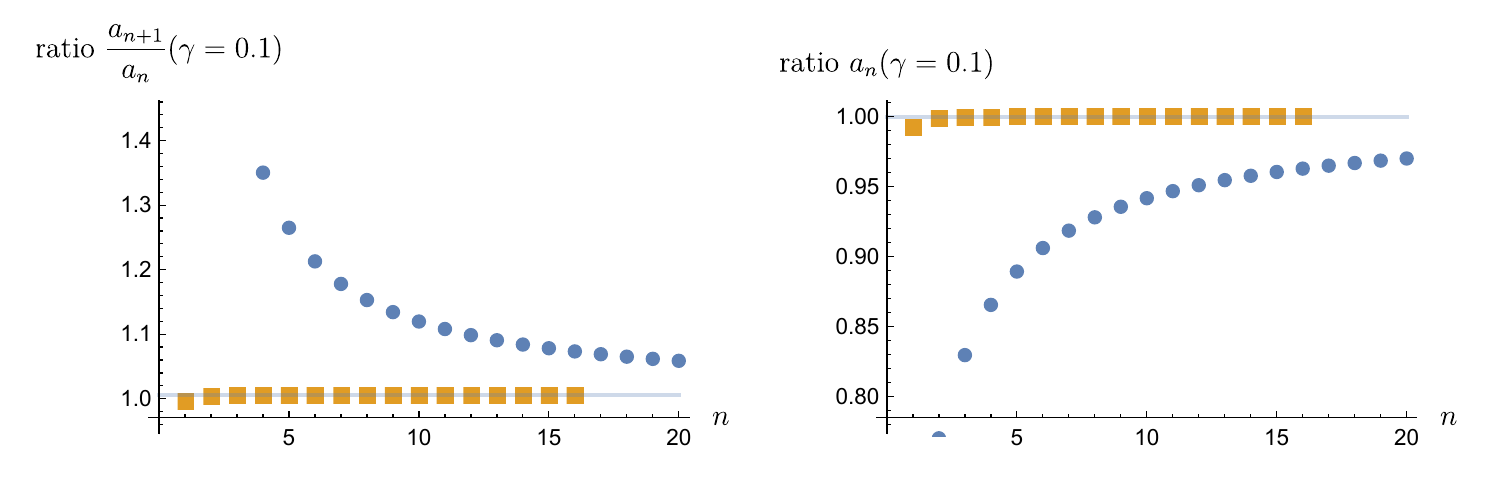}
    \caption{(Left) the left-hand side of \eqref{eq:an-ratio-test} at $\gamma=0.1$ for the first 20 coefficients (blue circles) and with a 4th-order Richardson (orange squares). The Richardson of the sequence converges rapidly to $\displaystyle \qty(\qty(\sqrt{1+\gamma^2}+1)/2)^2\approx 1.00499$. (Right) the ratio of $a_n(\gamma)$ to its large-order growth on the right-hand-side of \eqref{eq:leading-large}, at $\gamma=0.1$, for the first 20 coefficients (blue circles) and with a 4th-order Richardson (orange squares), which converges rapidly to 1.}
    \label{fig:an-growth-rate}
\end{figure}

The relation of the WKB approximation to resurgence can be understood by looking at the leading large-order growth of the $\gamma$ dependent exact expansion coefficients in \eqref{eq:an-gamma}. Straightforward ratio tests with only the first 20 coefficients reveal that the {\it leading} large-order behavior, as a function of $\gamma$, is:
\begin{eqnarray}
\frac{a_{n+1}(\gamma)}{a_n(\gamma)}\frac{(-1)}{\left(2n+\frac{3}{2}\right)\left(2n+\frac{1}{2}\right)}\sim \left(\frac{\sqrt{1+\gamma^2}+1}{2}\right)^2,
\qquad n\to\infty
\label{eq:an-ratio-test}
\end{eqnarray}
Therefore, the radius of convergence of the associated Borel transform is given by $2/(\sqrt{1+\gamma^2}+1)$, which matches the $\gamma$-dependent factor in the exponent of the WKB approximation (\ref{eq:wkb}). A more refined ratio-test analysis, combined with Richardson acceleration \cite{bender}, shows that the leading large order factorial growth rate is
\begin{eqnarray}
a_n(\gamma) \sim (-1)^n \frac{3\sqrt{\pi}}{4} \left(1+\gamma^2\right)^{5/4} \left(\frac{\sqrt{1+\gamma^2}+1}{2}\right)^{2n+3/2} \Gamma\left(2n+\tfrac{3}{2}\right)
\qquad, \quad n\to\infty
\label{eq:leading-large}
\end{eqnarray}
These ratio tests are shown in Figure \ref{fig:an-growth-rate}. 

\begin{figure}[t]
    \centering
    \includegraphics[width=\textwidth]{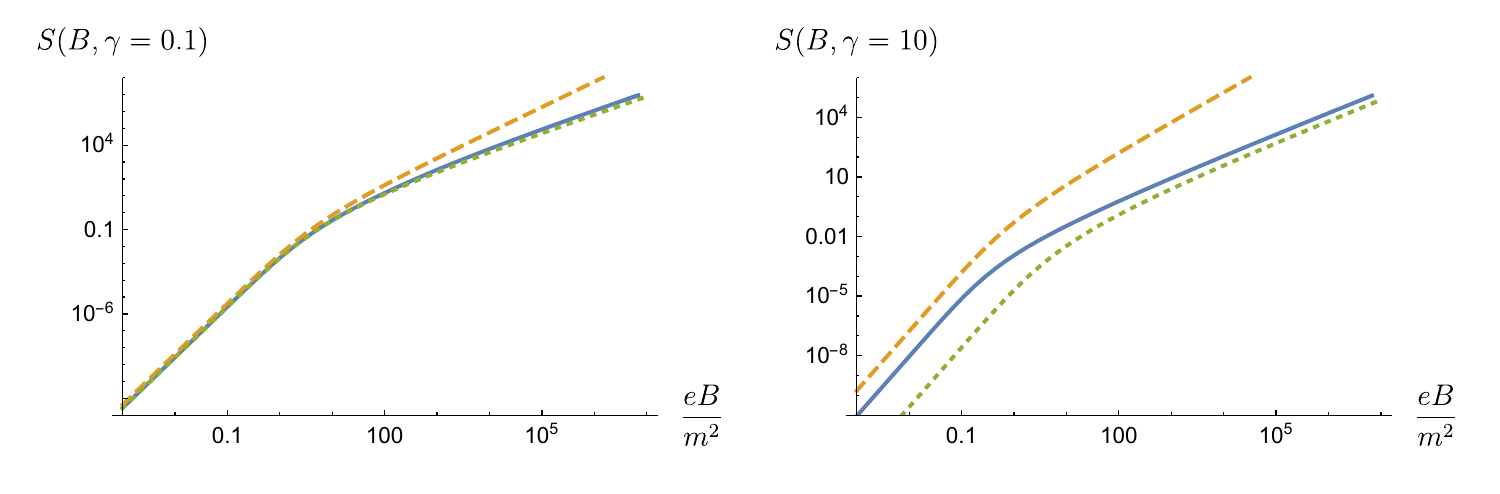}
    \caption{The exact effective action in (\ref{eq:exact2}) (blue, solid) compared with the WKB approximation for the magnetic field configuration (\ref{eq:magnetic-wkb}) (orange, dashed) and the locally constant field approximation (\ref{eq:borel-lcfa}) (green, dotted) for both small $\gamma$ (left) and large $\gamma$ (right). Both approximations are only accurate at small $\gamma$. Note the differences in structure and accuracy between this WKB approximation and the standard approximation for the electric field case in Figure \ref{fig:ImSapprox}, obtained by analytic continuation from \eqref{eq:magnetic-wkb}. These plots should also be compared with our resurgent extrapolations in Figure \ref{fig:magnetic-extrap}.}
    \label{fig:Sapprox}
\end{figure}

The WKB approximation for the inhomogeneous magnetic background field is obtained by replacing the exact expansion coefficients $a_n(\gamma)$ in \eqref{eq:weak-field-exp} by their leading behavior in \eqref{eq:leading-large}. Writing an integral representation for the gamma function factor, this yields:
\begin{equation}
    S_\text{WKB}(B,\lambda)\sim \lambda L^2T\,\frac{m^4}{4\pi^{3/2}}\qty(\frac{eB}{\pi m^2})^{5/2}\qty(1+\gamma^2)^{5/4}\int_0^\infty \dd{s}\exp\qty(-\frac{\pi m^2 s}{eB}\frac{2}{\sqrt{1+\gamma^2}+1})\frac{\sqrt{s}}{s^2+1}
    \label{eq:magnetic-wkb}
\end{equation}
The WKB prefactor $(1+\gamma^2)^{5/4}$ and exponent factor $2/(\sqrt{1+\gamma^2}+1)$ can both be seen directly in the leading large order growth \eqref{eq:leading-large}.
The result \eqref{eq:magnetic-wkb} should be compared with the magnetic LCFA expression in \eqref{eq:lcfa-magnetic}, in which neither of these correction factors appear, but which does incorporate the multi-instanton sum via the factor in \eqref{eq:he-partial}, as well as the subleading power law corrections in \eqref{eq:lcfa-b} via the hypergeometric U factor.
The geometric sum factor in \eqref{eq:magnetic-wkb} arises from the original perturbative sum. In Figure \ref{fig:Sapprox} we plot the exact effective action \eqref{eq:exact2} (blue solid curves), and compare it with the magnetic WKB expression \eqref{eq:magnetic-wkb} (orange dotted curves), as well as the LCFA expression for 
a magnetic background \eqref{eq:borel-lcfa} (green dotted curves), for both a small ($\gamma=0.1$, left) and large ($\gamma=10$, right) inhomogeneity parameter.
In the next Section we show how these approximations can be generalized to include all power-law and exponential corrections, yielding the full all-orders trans-series structure of the exact effective action \eqref{eq:exact2}.

\section{Trans-series Structure from Borel Analysis of the Exact Effective Action}
\label{sec:borel}

To probe in greater detail the resurgent properties of the QED effective action in an inhomogeneous background field, we turn to a Borel analysis of  the exact effective action in \eqref{eq:exact2}.
We first note that the Bose-like factor in \eqref{eq:exact2} may be expanded as a geometric series, yielding a geometric sum over multi-instanton terms, each of which is identical, up to rescaling. In the language of the worldline representation, these are the multiple windings of the same classical solution \cite{wli1,wli2}. This means that we can effectively regard the remaining factor of the integrand as a (modified) Borel transform function (recall $z$ defined in \eqref{eq:z}):\footnote{See also \cite{Hatsuda:2015owa}.}
\begin{align}
  \mathcal{B}(s,\gamma)=\dv{z}{s}\qty(1-\frac{4z}{3}+\frac{z^2}{5}\, _{2}F_{1}\qty(1,1;\tfrac{7}{2};z))-(s\to -s)-\Bigg(\frac{8}{3}s-2\gamma \qty(1+\frac{\gamma^2}{4}s^2)^{3/2}\,\text{arcsinh}\qty(\frac{\gamma}{2} s)\Bigg)
  \label{eq:exact-borel}
\end{align}
The singularities of this Borel transform encode the non-perturbative physics of this problem.

\subsection{New Borel Singularities of the Effective Action}
\label{sec:borel-sings}

From the hypergeometric function term in $\mathcal{B}(s,\gamma)$ we deduce branch point singularities when $z=1$:  
\begin{equation}
  z=1\quad\Rightarrow\quad \pm s_1=\pm\frac{2i}{\sqrt{1+\gamma^2}+1},\quad \pm s_2=\mp\frac{2i}{\sqrt{1+\gamma^2}-1}
  \label{eq:t1-t2}
\end{equation}
There is also  a complex conjugate pair of branch points from the subtraction term:
\begin{equation}
  \pm s_3=\pm\frac{2i}{\gamma}
  \label{eq:t3}
\end{equation}
After analytic continuation to the electric field case, these branch point singularities induce non-perturbative
imaginary contributions whose exponential parts are
\begin{equation}
  \exp\qty(-\frac{\pi m^2}{e E}\frac{2}{\sqrt{1+\gamma^2}+1})\quad,\quad   
  \exp\qty(-\frac{\pi m^2}{e E}\frac{2}{\sqrt{1+\gamma^2}-1}) \quad,\quad 
  \exp\qty(-\frac{\pi m^2}{eE}\frac{2}{\gamma})
  \label{eq:exps}
\end{equation}
\begin{figure}[htb]
  \centering
  \includegraphics[width=\textwidth]{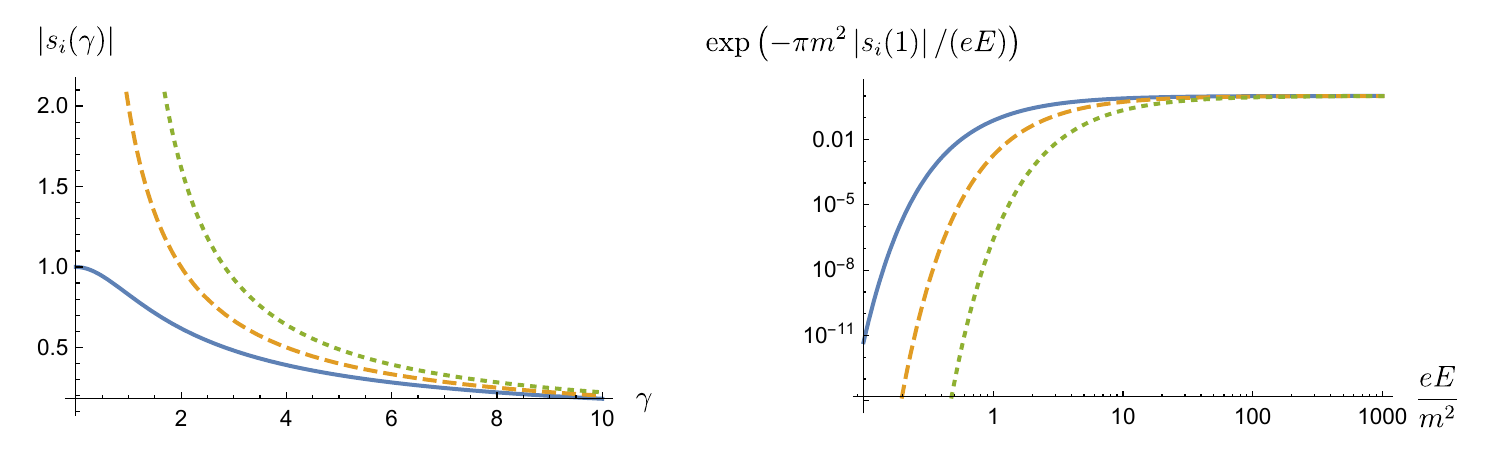}
  \caption{The left plot shows the modulus of the Borel singularity locations, $s_1$ (blue, solid), $s_2$ (green, dotted) and $s_3$ (orange, dashed) in (\ref{eq:t1-t2}) and (\ref{eq:t3}), as a function of the inhomogeneity parameter $\gamma$. For all $\gamma$ we see that $\qty|s_1|$ is the dominant singularity, then $\qty|s_3|$, and then  $\qty|s_2|$. See (\ref{eq:hierarchy}). The right plot shows the corresponding instanton factors in (\ref{eq:exps}), for $\gamma=1$, with the same color scheme as in the left-hand plot. }
  \label{fig:hierarchy}
\end{figure}
for $s_1$, $s_2$ and $s_3$, respectively. The first of these factors is recognized as the exponential factor in the standard WKB approximation (\ref{eq:wkb}), and which appears in the leading large order growth \eqref{eq:leading-large}. The other two exponential factors in \eqref{eq:exps} are novel saddle contributions whose origin can be understood using resurgent asymptotics, as explained below. In the constant field limit, $\gamma\to 0$, the contribution of the $s_1$ branch point reduces to the well-known Euler-Heisenberg constant field exponential factor $\exp\qty(-\frac{\pi m^2}{e E})$. On the other hand, the contribution of each of the new branch points $s_2$ and $s_3$ vanishes exponentially in the constant field limit, $\gamma\to 0$. However, as the inhomogeneity parameter $\gamma$ increases these new branch points become more and more physically relevant. The hierarchy of the singularities is determined by the fact that 
\begin{equation}
|s_1|< |s_3|<|s_2|\qquad \text{for all $\gamma$}
\label{eq:hierarchy}
\end{equation}
as shown in the left plot Figure \ref{fig:hierarchy}. The right plot in Figure \ref{fig:hierarchy} shows the hierarchy of the three non-perturbative factors in \eqref{eq:exps}, as functions of $\frac{eE}{m^2}$, for a chosen value of $\gamma=1$.

\subsection{Darboux's Theorem and the Large-Order Growth of the Perturbative Coefficients}
\label{sec:darboux}

By Darboux's theorem \cite{henrici}, the singularities of the Borel transform determine the large-order growth behavior of the coefficients of its expansion about $s=0$, and hence they determine the large-order behavior of the expansion coefficients of the weak magnetic field expansion of the effective action. These coefficients depend on $\gamma$, so we are in the situation of ``parametric resurgent asymptotics''. We can write this expansion about $s=0$ in terms of the perturbative coefficients $a_n(\gamma)$ in \eqref{eq:an-gamma0}-\eqref{eq:an-gamma} as
\begin{equation}
    \mathcal{B}(s,\gamma)=\sum_{n=0}^\infty \frac{a_n(\gamma)}{\Gamma(2n+4)\zeta(2n+4)} s^{2n+3}\label{Borel}
\end{equation}
Simple ratio tests (recall Figure \ref{fig:an-growth-rate}) confirm the expectation that the radius of convergence is equal to $\qty|s_1|$. Using the transformation properties of hypergeometric functions, we can express the exact Borel transform $\mathcal{B}(s,\gamma)$ in \eqref{eq:exact-borel} in a form that makes all its singularities completely explicit, not just the leading one:
\begin{equation}
\begin{aligned}
  \mathcal{B}(s,\gamma)&=\qty[\qty(1-\frac{s}{s_1})^{3/2}\qty(1-\frac{s}{s_2})^{3/2}\xi(s,\gamma)+\psi(s,\gamma)]-(s\to -s)\\
  &\quad +\qty(1-\frac{s}{s_3})^{3/2}\qty(1+\frac{s}{s_3})^{3/2}\zeta(s,\gamma)-\frac{8}{3}s
\end{aligned}
\label{eq:borel-sings}
\end{equation}
Here we have defined
\begin{align}
  \xi(s,\gamma)&=\pi\dv{\sqrt{z}}{s}\\
  \zeta(s,\gamma)&=2\gamma\,\text{arcsinh}\qty(\frac{\gamma}{2} s)\\
  \psi(s,\gamma)&=-2\dv{\sqrt{z}}{s}\qty(1-z)^{3/2}\arcsin\sqrt{1-z}
\end{align}
where we recall (\ref{eq:z}) that $z$ is a simple quadratic function of $s$: $z(s)=-is-\gamma^2 s^2/4$. Given this result it is straightforward to expand about each of the singularities:
\begin{align}
  \qty(1-\frac{s}{s_1})^{3/2}\qty(1-\frac{s}{s_2})^{3/2}\xi(s,\gamma)&:=\qty(1-\frac{s}{s_1})^{3/2}\sum_{n=0}^\infty b_n^{(1)}(\gamma)(s-s_1)^n
  \nonumber\\
  &=\qty(1-\frac{s}{s_1})^{3/2}b_0^{(1)}(\gamma)\qty[1+i\frac{\qty(1-\frac{3}{4}\gamma^2)}{\sqrt{1+\gamma^2}}\qty(s-s_1)-\frac{3}{2}\frac{\qty(1+\frac{1}{4}\gamma^2)^2}{\qty(1+\gamma^2)}(s-s_1)^2+\ldots]
  \label{eq:t1exp}\\
  \qty(1-\frac{s}{s_1})^{3/2}\qty(1-\frac{s}{s_2})^{3/2}\xi(s,\gamma)&:=\qty(1-\frac{s}{s_2})^{3/2}\sum_{n=0}^\infty b_n^{(2)}(\gamma)(s-s_2)^n
  \nonumber\\
  &=\qty(1-\frac{s}{s_2})^{3/2}b_0^{(2)}(\gamma)\qty[1-i\frac{\qty(1-\frac{3}{4}\gamma^2)}{\sqrt{1+\gamma^2}}\qty(s-s_2)-\frac{3}{2}\frac{\qty(1+\frac{1}{4}\gamma^2)^2}{\qty(1+\gamma^2)}(s-s_2)^2+\ldots]
    \label{eq:t2exp}\\
  \qty(1-\frac{s}{s_3})^{3/2}\qty(1+\frac{s}{s_3})^{3/2}\zeta(s,\gamma)&:=\qty(1-\frac{s}{s_3})^{3/2}\sum_{n=0}^\infty b_n^{(3)}(\gamma)(s-s_3)^n+\text{analytic}
  \nonumber\\
  &=\qty(1-\frac{s}{s_3})^{3/2}b_0^{(3)}(\gamma)\qty[1-\frac{3i}{4}\gamma(s-s_3)-\frac{3}{32}\gamma^2(s-s_3)^2+\ldots]+\text{analytic}
    \label{eq:t3exp}
\end{align}
with
\begin{align}
    b_0^{(1)}(\gamma)&=-\frac{i\pi}{2}\qty|s_1|^{3/2}(1+\gamma^2)^{5/4}\\
    b_0^{(2)}(\gamma)&=\frac{i\pi}{2}\qty|s_2|^{3/2}\qty(1+\gamma^2)^{5/4}\\
    b_0^{(3)}(\gamma)&=i\pi \qty|s_3|^{3/2}\gamma^{5/2}
\end{align}

These expansions uniquely define three sets of expansion coefficients, $b_n^{(1)}(\gamma)$, $b_n^{(2)}(\gamma)$ and $b_n^{(3)}(\gamma)$,  which describe the local behavior of the Borel transform in the neighborhood of the associated Borel singularity, $s_1$, $s_2$ or $s_3$, respectively. These coefficients can be generated recursively, or written as explicit sums. We note that 
\begin{equation}
    \frac{b_n^{(1)}(\gamma)}{b_0^{(1)}(\gamma)}=(-1)^n\frac{b_n^{(2)}(\gamma)}{b_0^{(2)}(\gamma)}\qquad \text{for all }n
    \label{eq:bn1-bn2}
\end{equation}
Furthermore, we note that the $b_n^{(3)}(\gamma)$ coefficients match the coefficients of the leading large $\gamma$ terms in $b_n^{(1)}(\gamma)$:
\begin{equation}
    \frac{1}{\gamma^n}\frac{b_n^{(3)}(\gamma)}{b_0^{(3)}(\gamma)}=\lim_{\gamma\to\infty}\qty[\frac{1}{\gamma^n}\frac{b_n^{(1)}(\gamma)}{b_0^{(1)}(\gamma)}]\label{eq:limit}
\end{equation}
\begin{figure}[t]
  \centering
  \includegraphics[width=\textwidth]{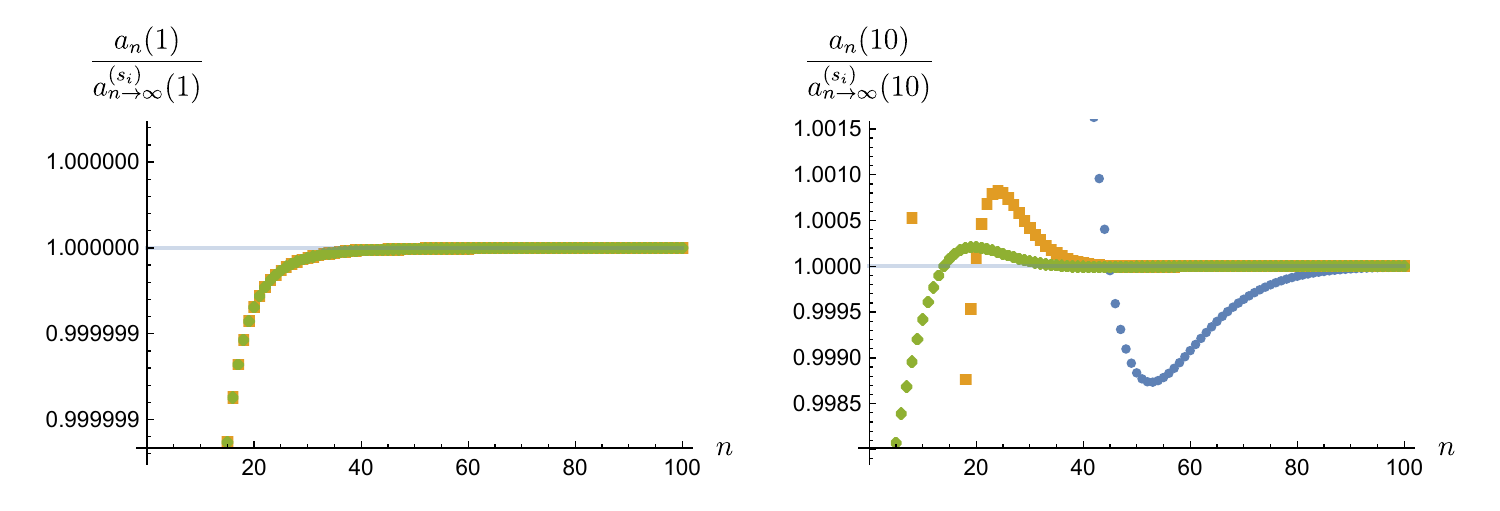}
  \caption{The ratio of the first 100 coefficients in the expansion of $\mathcal{B}(s,\gamma)$ about $s=0$, expressed in terms of $a_n(\gamma)$, to the leading ($k=0$) terms in the large-order growth predicted in (\ref{eq:an-gamma-large-order}). We display separately the contribution from just $s_1$ (blue circles), the sum of the contributions from $s_1$ and $s_3$ (orange squares), and the sum of the contributions from all three singularities (green diamonds), for both $\gamma=1$ (left) and $\gamma=10$ (right). At large $\gamma$, the sub-leading singularities make significant contributions and must be included for an accurate representation of the Borel transform function $\mathcal{B}(s,\gamma)$.} 
   \label{fig:an-gamma-ratio}
\end{figure}

Darboux's theorem predicts that the large-order growth of the coefficients of the expansion of the Borel transform about $s=0$ can be expressed in terms of the expansion coefficients about the function's singularities. The leading growth comes from the pair of leading singularities at $\pm s_1$:
\begin{align}
\frac{a_n(\gamma)}{\Gamma(2n+4)}
    &\sim \frac{3\sqrt{\pi}}{4}(-1)^n \frac{\Gamma\qty(2n+\tfrac{3}{2})}{\Gamma(2n+4)}\frac{\qty(1+\gamma^2)^{5/4}}{\qty|s_1|^{2n+3/2}} \zeta(2n+4)\nonumber\\
    &\quad\times\qty[1-\frac{5}{4}\frac{\qty(1-\tfrac{3}{4}\gamma^2)}{\sqrt{1+\gamma^2}}\frac{\qty|s_1|}{\qty(2n+\tfrac{1}{2})}+\frac{105}{32}\frac{\qty(1+\tfrac{1}{4}\gamma^2)^2}{(1+\gamma^2)}\frac{\qty|s_1|^2}{\qty(2n+\tfrac{1}{2})\qty(2n-\tfrac{1}{2})}+\ldots], \qquad n\to\infty\label{eq:BorelGrowth}
\end{align}
This result explains the leading behavior in \eqref{eq:leading-large}, including the overall Stokes constant. Furthermore, it generalizes the leading expression \eqref{eq:leading-large} by including the $\gamma$-dependent subleading power-law corrections, in which we recognize the appearance of the $b_n^{(1)}(\gamma)$ functions, coming from the expansion \eqref{eq:t1exp} of the modified Borel transform function $\mathcal{B}(s,\gamma)$ near the leading singularities at $\pm s_1$. 
Including the expansions \eqref{eq:t2exp}-\eqref{eq:t3exp} near the leading sub-leading singularities at $\pm s_2$ and $\pm s_3$, we deduce the complete all-orders large-order growth of the weak-field expansions coefficients $a_n(\gamma)$:
\begin{equation}
\begin{aligned}
a_n(\gamma)&\sim
  2\,\zeta(2n+4)
\sum_{\ell=0}^\infty(-1)^\ell\frac{\Gamma(2n+\tfrac{3}{2}-\ell)}{\Gamma\qty(-\tfrac{3}{2}-\ell)}\Bigg[\frac{b_\ell^{(1)}(\gamma)}{s_1^{2n+3-\ell}}+\frac{b_\ell^{(2)}(\gamma)}{s_2^{2n+3-\ell}}+\frac{b_\ell^{(3)}(\gamma)}{s_3^{2n+3-\ell}}\Bigg], \qquad n\to\infty
\end{aligned}
\label{eq:an-gamma-large-order}
\end{equation}
The pair of singularities at $\pm s_1$ is always dominant, because of the hierarchy \eqref{eq:hierarchy}, but for $\gamma \gtrsim 1$ the sub-leading singularities begin to contribute significantly. In Figure \ref{fig:an-gamma-ratio} this large-order behavior  is confirmed numerically. We generate the first 100 coefficients in the expansion of $\mathcal{B}(s,\gamma)$ around $s=0$, and plot the ratio of these exact coefficients to the $\ell=0$ terms of the large-order prediction in (\ref{eq:an-gamma-large-order}); (i) using just the dominant contribution involving $s_1$ (blue points), (ii) using the first two dominant contributions involving $s_1$ and $s_3$ (orange points), and (iii) using all three dominant contributions involving $s_1$, $s_3$ and $s_2$ (green points). As expected, for a small value $\gamma=1$ the leading singularity at $s_1$ dominates (note that for $\gamma=1$ we have $|s_1|\approx 0.828$, $|s_3|=2$ and $|s_2|\approx 4.828$). On the other hand, for a large value $\gamma=10$ the inclusion of all three singularities gives a much more accurate result (note that for $\gamma=10$ we have $|s_1|\approx 0.181$, $|s_3|\approx 0.2$ and $|s_2|\approx 0.221$).

This is a partial explanation for the disagreement at large $\gamma$ between the LCFA or the WKB approximation and the exact effective action. Recall Figures \ref{fig:ImSapprox} and \ref{fig:Sapprox}. Neither of these approximations includes any effect of the new higher Borel singularities at $\pm s_2$ and $\pm s_3$, and these become important for more inhomogeneous fields (large $\gamma$). We will see below that this discrepancy can be resolved using resurgent extrapolation methods.

\subsection{Resurgence for Inhomogeneous Background Fields}
\label{sec:resurgence}

Resurgence suggests that generically there are explicit relations between the large-order growth of the perturbative coefficients of an asymptotic expansion and the low-order coefficients of the fluctuations about neighboring non-perturbative saddle points \cite{berry-howls,Ecalle,costin-book,Marino:2012zq,Dorigoni:2014hea,gokce}. Here we explore this expectation for the QED effective action with an inhomogeneous electric field. The non-perturbative saddle points are defined by the singularities of the Borel transform, so the expansion of the Borel transform in the neighborhood of its singularities leads to expressions for the fluctuations about instanton terms in the imaginary part of the effective action.

We now translate the results from Darboux's theorem into resurgence relations in the physical, complex field strength plane, including all ``$k$-instanton'' contributions from a given singularity. 
Different ways of expressing the large-order growth of the perturbative coefficients lead to different representations of the effective action, highlighting different aspects of the resurgent structure. 

Consider first the contribution of the leading Borel singularity, $s_1$, to the large-order growth in (\ref{eq:an-gamma-large-order}):
\begin{align}
  S(B,\lambda)&\sim \lambda L^2T\,\frac{m^4}{3\pi^2}\sum_{n=0}^\infty \qty[2\zeta(2n+4)\sum_{\ell=0}^\infty (-1)^\ell\frac{\Gamma\qty(2n+\tfrac{3}{2}-\ell)}{\Gamma\qty(-\tfrac{3}{2}-\ell)}\frac{b_\ell^{(1)}(\gamma)}{s_1^{2n+3-\ell}}]\qty(\frac{eB}{\pi m^2})^{2n+4},\qquad eB\ll m^2
\end{align}
Using an integral representation for $\Gamma\qty(2n+\tfrac{3}{2}-\ell)$, we obtain
\begin{equation}
    S(B,\lambda)\sim \lambda L^2T\,\frac{2m^4}{3\pi^2}\qty(\frac{eB}{\pi m^2})^{\!4}\frac{i}{\qty|s_1|^3}\sum_{\ell=0}^\infty (-i\qty|s_1|)^\ell\frac{b_\ell^{(1)}(\gamma)}{\Gamma\qty(-\tfrac{3}{2}-\ell)}\int_0^\infty \dd{s}e^{-s}s^{1/2-\ell}\sum_{n=0}^\infty (-1)^n\zeta(2n+4)\qty(\frac{eBs}{\pi m^2 \qty|s_1|})^{2n}\label{eq:resurgence-start}
\end{equation}
After rescaling the Borel variable $s$, we recognize the sum inside the integral as the Borel transform of the Euler-Heisenberg effective Lagrangian:
\begin{equation}
    \begin{aligned}
        S(B,\lambda)&\sim -\lambda L^2T\,\frac{m^4}{8\pi^{1/2}}\qty(\frac{eB}{\pi m^2})^{\!5/2}\qty(1+\gamma^2)^{5/4}\int_0^\infty\frac{\dd{s}}{s^2}\exp\qty(-\frac{\pi m^2s}{eB}\frac{2}{\sqrt{1+\gamma^2}+1})\\
        &\quad\times\qty(\coth(\pi s)-\frac{1}{\pi s}-\frac{\pi s}{3})\frac{1}{\sqrt{s}}\qty[1-\frac{5}{4}\frac{\qty(1-\frac{3}{4}\gamma^2)}{\sqrt{1+\gamma^2}}\qty(\frac{eB}{\pi m^2s})+\frac{105}{32}\frac{\qty(1+\frac{1}{4}\gamma^2)^2}{(1+\gamma^2)}+\ldots]
    \end{aligned}\label{eq:exactborelsums1expanded}
\end{equation}
Different limits of this expression lead to the LCF approximation \eqref{eq:borel-lcfa} or to the WKB approximation \eqref{eq:magnetic-wkb}. Setting $\gamma=0$ we arrive at
\begin{equation}
    \begin{aligned}
        S_{\text{LCFA}}(B,\lambda)&\sim -\lambda L^2T\frac{m^4}{8\pi^{1/2}}\qty(\frac{eB}{\pi m^2})^{\!5/2}\int_0^\infty\frac{\dd{s}}{s^2}e^{-\pi m^2s/(eB)}\qty(\coth(\pi s)-\frac{1}{\pi s}-\frac{\pi s}{3})\\
        &\quad\times\frac{1}{\sqrt{s}}\qty[1-\frac{5}{4}\qty(\frac{eB}{\pi m^2s})+\frac{105}{32}\qty(\frac{eB}{\pi m^2 s})^2-\frac{1575}{128}\qty(\frac{eB}{\pi m^2s})^3+\ldots]
    \end{aligned}
\end{equation}
which agrees precisely with the LCFA expression \eqref{eq:borel-lcfa}. 
Furthermore, after analytic continuation to the background electric field, summing over the residues from all the multi-instanton Borel poles leads to the LCFA expression \eqref{eq:lcfa} for the imaginary part of the effective action.

The magnetic WKB approximation \eqref{eq:magnetic-wkb} arises from a different limit of \eqref{eq:exactborelsums1expanded}: we keep the $\gamma$ dependence, but take just the leading singularity of the $\frac{1}{s^2}\qty(\coth(\pi s)-\frac{1}{\pi s}-\frac{\pi s}{3})$ factor, as in \eqref{eq:he-partial}, and we take only the leading ($n=0$) fluctuation factor. This produces exactly the magnetic WKB approximation obtained above in \eqref{eq:magnetic-wkb}. When analytically continued to the electric field, the residue at the pole leads to the WKB expression \eqref{eq:wkb} for the imaginary part of the effective action.

Having seen how this works for the contribution from the leading singularity in the large order growth (\ref{eq:an-gamma-large-order}), it is straightforward to include the effect of the other two Borel singularities:
\begin{equation}
    \begin{aligned}
        S(B,\lambda)&\sim -\lambda L^2T\,\frac{m^4}{8\pi^{1/2}}\qty(\frac{eB}{\pi m^2})^{\!5/2}\qty(1+\gamma^2)^{5/4}\int_0^\infty\frac{\dd{s}}{s^2}\qty(\coth(\pi s)-\frac{1}{\pi s}-\frac{\pi s}{3})\\
        &\quad\times \Bigg[\exp\qty(-\frac{\pi m^2s}{eB}\frac{2}{\sqrt{1+\gamma^2}+1})\frac{1}{\sqrt{s}}\sum_{n=0}^\infty (-i)^n\frac{\Gamma\qty(-\tfrac{3}{2})}{\Gamma\qty(-\frac{3}{2}-n)}\frac{b_n^{(1)}(\gamma)}{b_0^{(1)}(\gamma)}\qty(\frac{eB}{\pi m^2 s})^n\\
        &\quad -2\exp\qty(-\frac{\pi m^2s}{eB}\frac{2}{\gamma})\qty(\frac{\gamma^2}{1+\gamma^2})^{5/4}\frac{1}{\sqrt{s}}\sum_{n=0}^\infty (-i)^n\frac{\Gamma\qty(-\tfrac{3}{2})}{\Gamma\qty(-\frac{3}{2}-n)}\frac{b_n^{(3)}(\gamma)}{b_0^{(3)}(\gamma)}\qty(\frac{eB}{\pi m^2 s})^n\\
        &\quad +\exp\qty(-\frac{\pi m^2s}{eB}\frac{2}{\sqrt{1+\gamma^2}-1})\frac{1}{\sqrt{s}}\sum_{n=0}^\infty i^n\frac{\Gamma\qty(-\tfrac{3}{2})}{\Gamma\qty(-\frac{3}{2}-n)}\frac{b_n^{(2)}(\gamma)}{b_0^{(2)}(\gamma)}\qty(\frac{eB}{\pi m^2 s})^n\Bigg]
    \end{aligned}\label{eq:exactresurgence}
\end{equation}
This representation \eqref{eq:exactresurgence} for the exact effective action is equivalent to (\ref{eq:exact2}), but has certain physical advantages:
\begin{itemize}
    \item 
    Equation \eqref{eq:exactresurgence} makes more explicit the relationship to the Euler-Heisenberg effective action, by the inclusion of the factor $\frac{1}{s^2}\qty(\coth(\pi s)-\frac{1}{\pi s}-\frac{\pi s}{3})$, which is the Euler-Heisenberg Borel transform function in \eqref{eq:l1b}.
    \item
    It also makes clear the effect of the three fundamental Borel singularities \eqref{eq:t1-t2}-\eqref{eq:t3}, whose magnitudes appear in the exponents.
    \item
    The sums over $n$ generate the asymptotic series of fluctuations about each singularity.
    \item
    Upon analytic continuation to the electric field case, the imaginary part is given by half the sum of the residues of the poles from the Euler-Heisenberg Borel transform
\begin{equation}
    \begin{aligned}
        \Im S(E,\tau)&\sim L^3\tau\,\frac{m^4}{8\pi^{1/2}}\qty(\frac{eE}{\pi m^2})^{\!5/2}\qty(1+\gamma^2)^{5/4}\sum_{k=1}^\infty\frac{1}{k^{5/2}}\Bigg[e^{-k\pi m^2\qty|s_1|/(eE)}\sum_{n=0}^\infty (-i)^n\frac{\Gamma\qty(-\tfrac{3}{2})}{\Gamma\qty(-\frac{3}{2}-n)}\frac{b_n^{(1)}(\gamma)}{b_0^{(1)}(\gamma)}\qty(\frac{eE}{k\pi m^2})^n\\
        &\quad -2\qty(\frac{\gamma^2}{1+\gamma^2})^{5/4}e^{-k\pi m^2\qty|s_3|/(eE)}\sum_{n=0}^\infty (-i)^n\frac{\Gamma\qty(-\tfrac{3}{2})}{\Gamma\qty(-\frac{3}{2}-n)}\frac{b_n^{(3)}(\gamma)}{b_0^{(3)}(\gamma)}\qty(\frac{eE}{k\pi m^2})^n\\
        &\quad +e^{-k\pi m^2\qty|s_2|/(eE)}\sum_{n=0}^\infty i^n\frac{\Gamma\qty(-\tfrac{3}{2})}{\Gamma\qty(-\frac{3}{2}-n)}\frac{b_n^{(2)}(\gamma)}{b_0^{(2)}(\gamma)}\qty(\frac{eE}{k\pi m^2})^n\Bigg]
    \end{aligned}\label{eq:Imexactresurgence}
\end{equation}
\end{itemize}  
Since the repeated instanton contributions are identical up to a re-scaling, the instanton sums can be written exactly in terms of polylogarithms
\begin{equation}
    \begin{aligned}
        \Im S(E,\tau)&\sim L^3 \tau\,\frac{m^4}{8\pi^{1/2}}\qty(1+\gamma^2)^{5/4}\sum_{n=0}^\infty (-i)^n\frac{\Gamma\qty(-\frac{3}{2})}{\Gamma\qty(-\frac{3}{2}-n)}\qty(\frac{eE}{\pi m^2})^{\!n+5/2}\Bigg[\frac{b_n^{(1)}(\gamma)}{b_0^{(1)}(\gamma)}\text{Li}_{n+5/2}\qty(e^{-\pi m^2\qty|s_1|/(eE)})\\
        &\quad -2\qty(\frac{\gamma^2}{1+\gamma^2})^{\!5/4}\frac{b_n^{(3)}(\gamma)}{b_0^{(3)}(\gamma)}\text{Li}_{n+5/2}\qty(e^{-\pi m^2\qty|s_3|/(eE)})+(-1)^n \frac{b_n^{(2)}(\gamma)}{b_0^{(2)}(\gamma)}\text{Li}_{n+5/2}\qty(e^{-\pi m^2\qty|s_3|/(eE)})\Bigg]
    \end{aligned}\label{eq:Imexactpolylogs}
\end{equation}
This is the exact multi-instanton expression for the imaginary part of the effective action for the electric field background (\ref{eq:E}), including the effect of all three independent Borel singularities $s_1$, $s_2$ and $s_3$. This result demonstrates that the coefficients in the perturbative sector, encode all the information about all the non-perturbative physics of the effective action. 

In Appendix A we show that this structure may be extracted from the expansion coefficients themselves, not just from the exact integral representation \eqref{eq:exact2} of the effective action. Another interesting representation of the effective action in terms of incomplete gamma functions is derived in the Appendix B.

\section{Resurgent Extrapolation of the Effective Action with Inhomogeneous Background Fields}
\label{sec:extrapolation}

Having shown that the perturbative sector encodes a wealth of resurgent non-perturbative information, as a function of the inhomogeneity parameter $\gamma$, we now turn to a more general approach that does not rely on knowledge of the explicit Borel transform function \eqref{eq:exact-borel}, but just on knowledge of a {\it finite number} of the perturbative weak magnetic field expansion coefficients $a_n(\gamma)$ in \eqref{eq:an-gamma0}-\eqref{eq:an-gamma}. We ask how much non-perturbative information can be decoded from a {\it finite} number of terms in the perturbative sector. An important practical consequence of resurgence is that it suggests improved methods of extrapolation from one parametric regime to another. This is ultimately based on the idea that for a resurgent function the expansions about different parametric regimes are not independent of one another, and one can develop extrapolation methods that take advantage of this structure. The surprising effectiveness of this approach in the constant field Euler-Heisenberg case at both one-loop \cite{florio} and two-loop \cite{Dunne:2021acr}, where the Borel singularities  are poles and branch points, respectively, motivates our analysis of the inhomogeneous background field problem, where further new Borel structure emerges. The more precisely the analytic structure of the Borel transform is understood, the more precisely the effective action can be analytically continued from one parametric region to another, in terms of the physical parameters. The basic analytic tools are simple to implement: Pad\'e approximants and conformal and uniformizing maps \cite{Costin:2020hwg,Costin:2020pcj,Costin:2021bay}.

In principle, the method described in this Section does not rely on the fact that the background field has the special property that the spectral information is soluble, except that this property provides more precise access to concrete comparisons. We anticipate that these methods should apply to more general inhomogeneous background fields, provided a modest number of perturbative coefficients can be computed for a given background field. 

In practical terms the technical problem is the following: 
\begin{quote}
    Given a {\it finite} number of terms of an asymptotic weak field expansion, where the coefficients depend also on inhomogeneity parameters (in our concrete example, this is the dependence on the inhomogeneity parameter $\gamma$), how much information can we extract about the singularity structure of the associated Borel transform, and hence information about the associated non-perturbative physics?
\end{quote}

Consider an $N$-term truncated weak magnetic field perturbative expansion of the effective action for an inhomogeneous magnetic field: 
\begin{eqnarray}
  S_N(B,\lambda)\sim \lambda L^2T\,\frac{m^4}{3\pi^2}\sum_{n=0}^{N-1}a_n(\gamma)\qty(\frac{eB}{\pi m^2})^{2n+4},\qquad eB\ll m^2
  \label{eq:weak-field-truncated}
\end{eqnarray}

Recall that from a small (approx. 20) finite number of terms we can deduce numerically the leading large-order growth \eqref{eq:leading-large}, which we write more compactly as
\begin{eqnarray}
a_n(\gamma) \sim (-1)^n \frac{3\sqrt{\pi}}{4} \left(1+\gamma^2\right)^{5/4} \frac{\Gamma\left(2n+\frac{3}{2}\right)}{|s_1|^{2n+3/2}}
\label{eq:leading-large2}
\end{eqnarray}
where we recall (\ref{eq:t1-t2}) that $s_1(\gamma)=2i/(\sqrt{1+\gamma^2}+1)$. The important pieces of information here are: (i) the power law $1/|s_1|^{2n+3/2}$, which determines the location of the leading Borel singularity; (ii) the gamma function offset $\frac{3}{2}$ which is associated with the nature of the leading Borel singularity; (iii) the overall Stokes constant, which determines the strength of the leading Borel singularity.

Since the weak field expansion of the effective action is asymptotic, of the generic ``factorial over power" form common in many physics applications \cite{dingle,LeGuillou:1990nq}, it is more accurate to extrapolate in the Borel plane than in the physical domain \cite{Costin:2020hwg}. Therefore we need to determine as accurately as possible the singularity structure in the Borel plane. This makes sense physically, because this Borel singularity structure represents the relevant non-perturbative physics. Pad\'e approximants are powerful, and easy to implement, tools to extract singularity information from finite-order expansions, and also to extrapolate beyond the radius of convergence \cite{baker,bender}. 

However, there are further considerations that imply that it is better to apply conformal map methods {\it before} using Pad\'e approximants. The numerical tests in the previous section suggest that the leading Borel singularities are branch points, rather than pole singularities. Furthermore, the physics of the problem, and our exact knowledge of the Euler-Heisenberg effective action for the constant background field, suggest that we should expect integer multiples of the leading Borel singularities (``multi-instantons", and possibly also other independent higher singularities). These last two items mean that we should not make Pad\'e approximants directly in the Borel plane, but we should first use a suitable conformal map, and {\it then} apply Pad\'e approximants inside the conformal disk. This Pad\'e-Conformal-Borel procedure can be shown to be significantly more accurate because it resolves more accurately the effect of the leading Borel singularities (which is the dominant effect), and furthermore because it includes more accurately the effect of Borel singularities beyond the leading ones \cite{Costin:2020hwg,Costin:2020pcj,Costin:2021bay}.

Given these considerations, we define the following truncated Borel transform function:
\begin{equation}
  \mathcal{B}_N(s,\gamma)=\sum_{n=0}^{N-1}\frac{a_n(\gamma)}{(2n+1)!} (|s_1|\, s)^{2n+2}\label{betterBorel}
\end{equation}
We have normalized the Borel variable by the magnitude $|s_1|$ of the leading singularities, which places the leading Borel singularities at $s=\pm i$. The extrapolated approximate effective action is then obtained by the inverse Borel transform:
\begin{align}
  S_N(B,\lambda)&=\lambda L^2T\, \frac{m^4}{3\pi^2}\qty(\frac{eB}{\pi m^2})^{\!2}\int_0^\infty \frac{\dd{s}}{s} e^{-\pi m^2\qty|s_1|s/(eB)}\mathcal{B}_N(s,\gamma)
  \label{IBPaction}
\end{align}
Our particular choice of Borel transform \eqref{betterBorel} and its associated inverse Borel transform integral \eqref{IBPaction}
is made in order to simplify the numerical analysis of the expected strong $B$ field behavior. We chose to isolate a factor of $s^{-1}$ from the Borel transform in the Borel integral \eqref{IBPaction}.
Correspondingly, we chose the power of $s$ in our Borel function
\eqref{betterBorel} to be ($2n+2$), rather than ($2n+1$).
Here we have used the well-known arbitrariness of choosing a suitable factorial factor to divide by when defining the Borel transform. The reason for doing this is that on physical grounds we know that in the strong field limit, $eB\gg m^2$, the effective action has logarithmic behavior $S\sim B^2 \ln (B)$ \cite{weisskopf,ritus1,ritus2,dunne-kogan}. The large $B$ limit of the effective action is determined by the large $s$ behavior of the Borel transform. Therefore, $S\sim B^2 \ln (B)$ behavior can be achieved most easily by the Borel integral \eqref{IBPaction} with a factor of $s^{-1}$ separated out, and with $\mathcal{B}_N(s,\gamma)\sim \text{constant}$ as $s\to\infty$. 

The truncated Borel transform in \eqref{betterBorel} is just a polynomial, so we need some method to explore the singularity structure of the function that it is trying to approximate as the order $N$ becomes large. Pad\'e \cite{baker,bender} is a remarkably efficient, and extremely simple, way to do this (although with low resolution).
We use Pad\'e approximation to represent the Borel transform as a ratio of polynomials\footnote{Note that the highest power of $s$ in \eqref{betterBorel} is $s^{2N}$.}
\begin{equation}
  \mathcal{P}^{N}_N\qty[\mathcal{B}_N](s,\gamma)=\frac{P_{N}(s,\gamma)}{Q_N(s,\gamma)}=\mathcal{B}_N(s,\gamma)+\mathcal{O}\qty(s^{2N+1})
  \label{eq:direct-pade}
\end{equation}
\begin{figure}[htb]
  \centering
  \includegraphics[width=\textwidth]{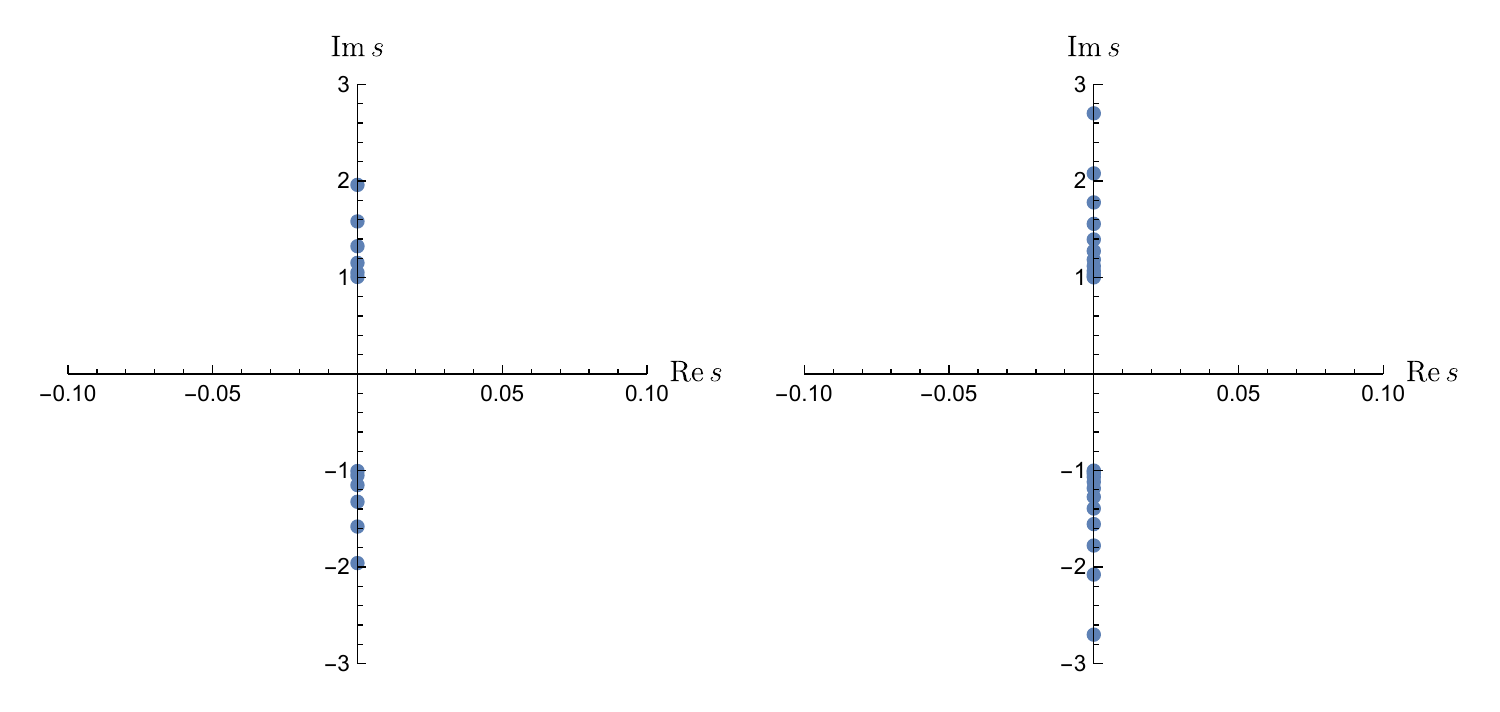}
  \caption{The Pad\'e poles of the finite-order Pad\'e-Borel transform \eqref{eq:direct-pade} for $N=15$ (left) and $N=30$ (right). The Pad\'e poles are attempting to form a pair of conjugate branch cuts starting at $\pm i$, in agreement with our exact knowledge of the Borel transform.}
  \label{fig:borel-poles}
\end{figure}
We primarily work with the diagonal $[N,N]$ Pad\'e approximant, but it is useful also to consider other near-diagonal Pad\'e approximants to help with identifying spurious poles.
The poles of the Pad\'e denominator $Q_N(s, \gamma)$ provide approximate information about the singularities of the actual Borel transform function. See Figure \ref{fig:borel-poles}. We observe Pad\'e poles accumulating into branch cuts at branch point locations $\pm i$, in agreement with what we know from the exact analytic structure of the exact Borel transform function studied in Section \ref{sec:borel}, and with the ratio tests of the previous subsection.

However, Figure \ref{fig:borel-poles} does not reveal anything about the other higher Borel singularities, at $s_2$ or $s_3$ in \eqref{eq:t1-t2}-\eqref{eq:t3}, or about multi-instanton integer repetitions of $s_1$, $s_2$ and $s_3$. This is because Pad\'e represents a branch cut emanating from the leading singularity as a line of poles (and interlaced zeros) that accumulate to the branch point location.\footnote{There is a physical interpretation of Pad\'e approximants in terms of two-dimensional electrostatics \cite{stahl,saff,Costin:2021bay}. This also explains why conformal maps are so useful, as they are a natural way to analyze two-dimensional electrostatics problems.} This means that genuine collinear singularities are hidden amongst these poles accumulating to the leading singularity. The conformal map provides a simple resolution of this shortcoming of Pad\'e, because it separates these higher singularities so that they can be resolved. As a bonus, in the process it also provides higher resolution near the branch points and branch cuts.
Therefore, a better strategy is to make a suitable conformal map {\it before} making the Pad\'e approximation. 
This generic improvement in the precision of the extrapolated Borel transform, especially near branch points and cuts, results in a significantly more accurate extrapolation in terms of the physical variable.

Based on the Pad\'e pole structure in Figure \ref{fig:borel-poles} for the direct  Pad\'e-Borel approximant in \eqref{eq:direct-pade}, it is natural to choose the following  2-cut conformal map which maps the doubly-cut Borel $s$ plane into the unit disk in the conformal $w$ plane:
\begin{equation}
  s= \frac{2w}{1-w^2}\qquad \longleftrightarrow\qquad w=\frac{s}{\sqrt{1+s^2}+1}
  \label{eq:cmap}
\end{equation}
This map takes $s=0$ to $w=0$, and $s=\pm i$ to $w=\pm i$, and maps the sides of the cuts to portions of the unit circle, $|w|=1$, the boundary of the unit disk.
The strategy of the Pad\'e-Conformal-Borel approximation is simple:
\begin{enumerate}
\item 
Re-expand $\mathcal B_N\left(\frac{2w}{1-w^2},\gamma\right)$ to the same order $2N$ (this is optimal \cite{Costin:2020pcj}).
\item
Make a Pad\'e approximant in the $w$ variable of the resulting expansion.
\item
Map back to the original Borel plane with the inverse map.
\end{enumerate}
In equations this reads:
\begin{align}
  \mathcal{B}_N(w,\gamma)&=\sum_{n=0}^{N-1}\frac{a_n(\gamma)}{\Gamma(2n+2)}\qty(\qty|s_1|\frac{2w}{1-w^2})^{2n+2}\quad \xrightarrow{{\rm re-expand}} \quad
  \mathcal{CB}_N(w,\gamma): = \sum_{n=0}^{2N}\alpha_n(\gamma)w^{n}\\
  \mathcal{P}^N_N\qty[\mathcal{CB}_N](w,\gamma)&:=\frac{\mathfrak{P}_N(w,\gamma)}{\mathfrak{Q}_N(w,\gamma)}=\sum_{n=0}^{2N}\alpha_n(\gamma)w^n+\mathcal{O}\qty(w^{2N+1})
\end{align}
We will see below that we obtain remarkably accurate numerical extrapolations and analytic continuations with as few as $15$ input terms. In Appendix C  we illustrate in detail how and why the Conformal-Pad\'e-Borel method resolves higher collinear Borel singularities, which ordinary Pad\'e-Borel cannot do. 

\begin{figure}[tb]
  \centering
  \includegraphics[width=\textwidth]{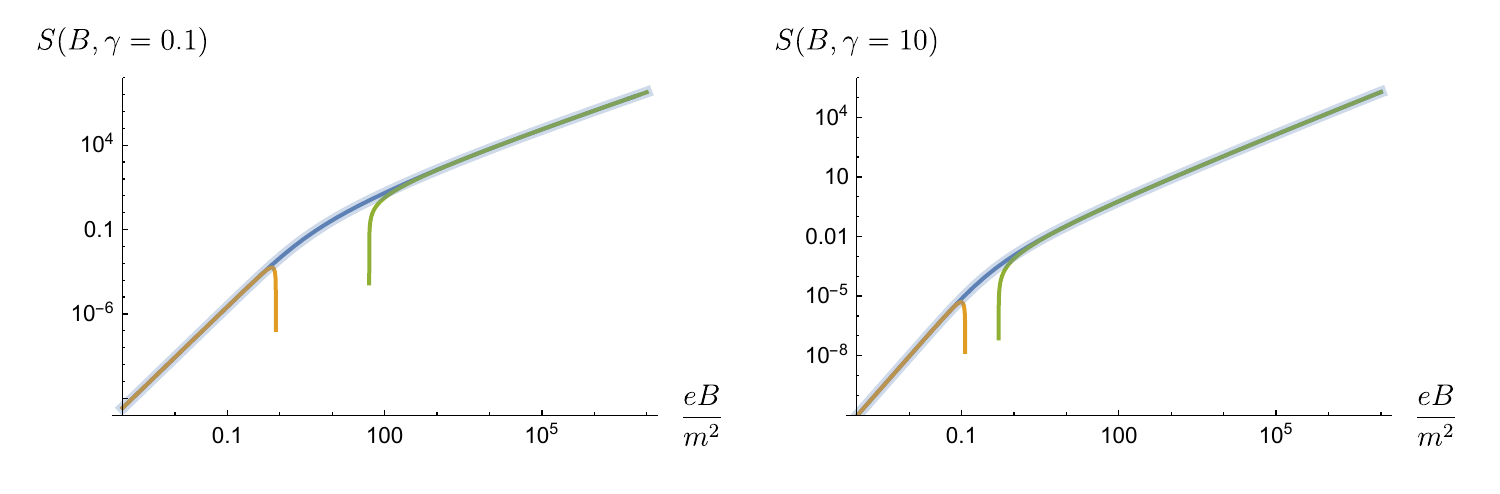}
    \caption{The exact effective action \eqref{eq:exact2} (blue) compared to truncating the weak magnetic field expansion at $N=15$ terms \eqref{eq:weak-field-truncated} (orange), the leading strong-field behavior \eqref{eq:large-b} (green), and the Pad\'e-Conformal-Borel approximation \eqref{eq:PCB-magnetic} using $N=15$ perturbative coefficients  (translucent blue band) at both small $\gamma$ (left) and large $\gamma$ (right). The Pad\'e-Conformal-Borel summation provides a high accuracy extrapolation from the weak-field to strong-field regime across ten orders of magnitude, and for both small and large inhomogeneity parameters $\gamma$. Contrast with the failure of both the LCF and WKB approximations for large inhomogeneity, as in Figure \ref{fig:Sapprox}. }
    \label{fig:magnetic-extrap}
\end{figure}

The final step is to apply the inverse conformal map in \eqref{eq:cmap} in order to return to the Borel plane
\begin{equation}
  \mathcal{P}^N_N[\mathcal{CB}_N](s,\gamma):=\mathcal{P}^N_N\qty[\mathcal{CB}_N]\qty(w=\frac{s}{\sqrt{1+s^2}+1},\gamma)
\end{equation}
Therefore, the final expression for our extrapolated ``Pad\'e-Conformal-Borel'' approximation to the effective action is
\begin{equation}
  \mathcal{CS}[S_N](B,\lambda)=\lambda L^2 T\frac{m^4}{3\pi^2}\qty(\frac{eB}{\pi m^2})^2\int_0^\infty \frac{\dd{s}}{s} e^{-\pi m^2\qty|s_1|s/(eB)}\mathcal{P}^N_N\qty[\mathcal{CB}_N](s,\gamma)\label{eq:PCB-magnetic}
\end{equation}
Plots of this summation are shown in Figure \ref{fig:magnetic-extrap} for $N=15$, in which we obtain a high precision extrapolation over ten orders of magnitude for both $\gamma$ small and large. We compare also with the logarithmic large $B$ behavior of the exact Borel integral \eqref{eq:exact2} (here $A$ is the Glaisher-Kinkelin constant):
\begin{equation}
    S(B,\lambda)\sim \lambda L^2T\,\frac{m^4}{3\pi^2}\cdot\frac{1}{3}\qty(\frac{eB}{m^2})^2\qty[\ln\qty(\frac{eB}{m^2})+\ln(2\gamma)-12\ln A+\frac{2}{3}],\qquad eB\gg m^2\label{eq:large-b}
\end{equation}
As a technical comment, we note that the logarithmic strong field behavior in \eqref{eq:large-b} arises from the following large $s$ behavior of the exact Borel transform function in \eqref{eq:exact-borel}:
\begin{equation}
    \mathcal{B}(s,\gamma)\sim 2s\ln(\gamma s)-\frac{2}{3}s
    \quad,\qquad s\to +\infty 
    \label{eq:large-s}
\end{equation}
This $\mathcal{B}(s, \gamma) \sim s\ln s$ behavior can be converted to the simpler $\mathcal{B}(s, \gamma) \sim s^{-1}$ behavior by integrating-by-parts twice.\footnote{This double integration-by-parts has an additional benefit of converting the $1/(e^{\pi m^2 s/(eB)}-1)$ factor in the integral representation \eqref{eq:exact2} into a factor ${\rm Li}_2\left(e^{-\pi m^2 s/(eB)}\right)$, which directly encodes the multi-instanton sum.} 
This then matches our chosen form of the Borel transform and integral in \eqref{betterBorel}-\eqref{IBPaction}.
In general the choice of extracting a factor of $s^{-1}$ from the Borel transform in \eqref{betterBorel}-\eqref{IBPaction} is because it is {\it numerically} easier to extrapolate from a small $s$ Taylor series to large $s$ power law behavior than to extrapolate from a small $s$ Taylor series to large $s$ logarithmic behavior. This is analogous to considerations of subtractions in dispersion relations, and makes our subsequent combination of conformal maps and Pad\'e approximants numerically consistent with the expected logarithmic strong field behavior of the effective action.

\begin{figure}[bt]
  \centering
  \includegraphics[width=\textwidth]{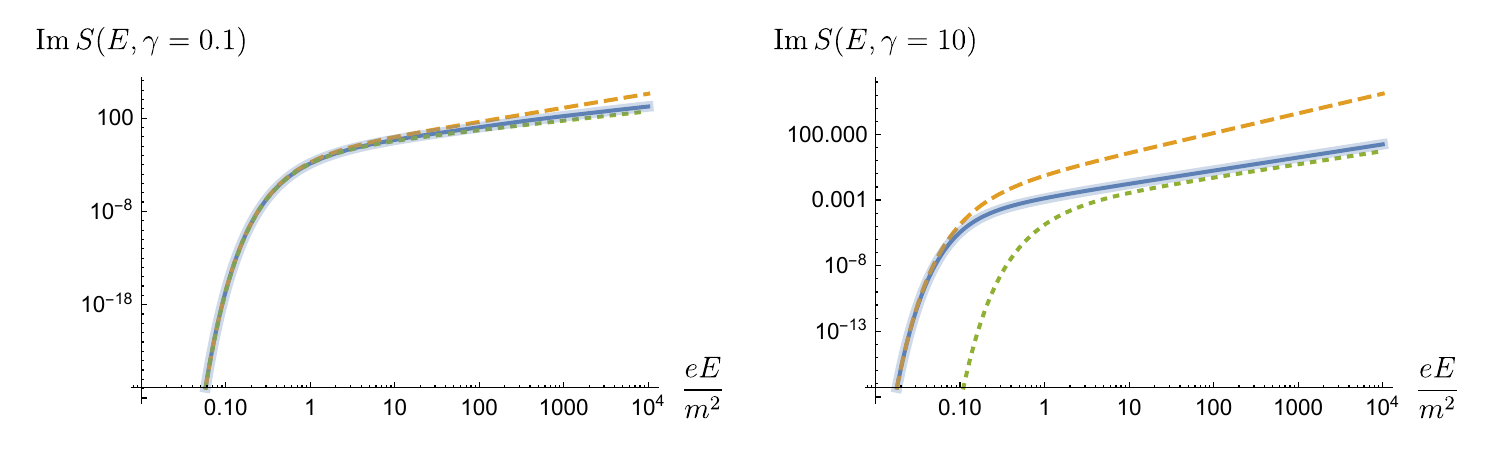}
  \caption{The exact imaginary part of the effective action in \eqref{eq:exact2} after analytically continuing to the electric field background \eqref{eq:E} (blue, solid) compared with the standard WKB approximation \eqref{eq:wkb} (orange, dashed), the locally constant field approximation \eqref{eq:lcfa} (green, dotted), and the lateral Pad\'e-Conformal-Borel approximation using $N=15$ perturbative coefficients \eqref{eq:PCB-electric} (translucent blue band) at both small $\gamma$ (left) and large $\gamma$ (right). The lateral summation was taken at $0^\pm=\pm 0.01$. The Pad\'e-Conformal-Borel summation provides a high accuracy extrapolation from the weak magnetic field regime to the electric field regime across six orders of magnitude and across a range of $\gamma$. Contrast with the failure of both the LCF and WKB approximations for large inhomogeneity, as in Figure \ref{fig:ImSapprox}.}
  \label{fig:PCB-electric}
\end{figure}

We can also analytically continue from a magnetic field background to an electric background, and define a lateral Borel sum
\begin{equation}
  \mathcal{CS}_\theta[S_N](E,\tau)=-L^3 \tau\frac{m^4}{3\pi^2}\qty(\frac{eE}{\pi m^2})^2\int_0^{\infty e^{i\theta}} \frac{\dd{s}}{s} e^{-\pi m^2\qty|s_1|s/(eE)}\mathcal{P}^N_N\qty[\mathcal{CB}_N](-is,\gamma)
\end{equation}
The imaginary part of the effective action is then given by
\begin{equation}
  \Im\mathcal{CS}[S_N](E,\tau)=\frac{1}{2}\Big(\mathcal{CS}_{0^+}[S_N](E,\tau)-\mathcal{CS}_{0^-}[S_N](E,\tau)\Big)\label{eq:PCB-electric}
\end{equation}
The result is shown in Figure \ref{fig:PCB-electric} for $N=15$ (with $0^\pm=\pm 0.01$). These extrapolations provide high precision across six orders of magnitude, and are a significant improvement over both the LCF approximation and the WKB approximation: compare with Figure \ref{fig:ImSapprox}. We emphasize that the extrapolations and analytic continuations in Figures \ref{fig:magnetic-extrap} and \ref{fig:PCB-electric} only used a modest number ($N=15$) of perturbative coefficients.

\section{Conclusions and Outlook}

We have shown that the one-loop QED effective action in an inhomogeneous background field has a resurgent asymptotic weak field expansion which encodes a wealth of non-perturbative information. The Borel singularity structure differs from that of the familiar Euler-Heisenberg constant background field case in two important ways: (i) there are two new types of Borel singularities whose effect vanishes in the constant field limit but which become significant for strongly inhomogeneous fields; (ii) the Borel singularities are branch points rather than poles, and have asymptotic fluctuation series multiplying the various instanton sectors, with these fluctuations being resurgently related to the perturbative expansion coefficients. Analogous statements apply to the gradient expansion itself, as can be seen by rescaling the double expansion to express it as a sum over derivatives, with coefficients that are functions of the inhomogeneity parameter $\gamma$. 

We also demonstrate that with a relatively modest amount of perturbative information one can reconstruct detailed non-perturbative information, extrapolating accurately from the weak magnetic field regime to the strong magnetic field regime, and also analytically continuing from the weak magnetic field regime to the electric field regime, recovering the exponentially small imaginary part associated with vacuum pair production. This motivates a potential new approach to these problems with more general inhomogeneities and at higher-loop order, where exact integral representations of the effective action are not known, but from which perturbative expansions may nevertheless be feasible. Our results show that such a perturbative expansion may encode relevant non-perturbative information even in the strongly inhomogeneous regime. Surprisingly, at one loop order we observed that this resurgent extrapolation approach is significantly more accurate than the locally constant field approximation and the WKB approximation, especially when the background field is strongly inhomogeneous.

\vskip 1cm
 This material is based upon work supported by the U.S. Department of Energy, Office of Science, Office of High Energy Physics under Award Number DE-SC0010339, and by a Graduate Research Fellowship Award from the National Science Foundation under Grant DEH-1747453.

 \section{Appendix A: Resurgence of the gradient expansion to all orders}
 \label{sec:appA}
 
 In this Appendix we show that the full non-perturbative structure may alternatively be reconstructed directly from the perturbative expansion (\ref{eq:weak-field-exp}), rather than from the exact large-order behavior in (\ref{eq:an-gamma-large-order}). Since the inhomogeneous background magnetic field (\ref{eq:B}) depends on two parameters, $B$ and $\lambda$, it is natural to expand the exact effective action as a double expansion in powers of the weak field $B$ and in powers of $\frac{1}{\lambda}$:\footnote{Because of symmetry, it is in fact an expansion in powers of $B^2$ and in powers of $1/\lambda^2$.}
\begin{eqnarray}
 S(B,\lambda)&\sim& -\lambda L^2T\frac{m^4}{8\pi^{3/2}}
  \sum_{j=0}^\infty 
  \left(\frac{1}{m\lambda}\right)^{2j}\sum_{n=0}^\infty c_n^{(j)}\qty(\frac{eB}{\pi m^2})^{2n+2}
  \label{eq:double}
\end{eqnarray}
The LCFA expansion \eqref{LCFAweakfield} corresponds to the leading order, $j=0$, of the gradient expansion (for which we recall that the $n=0$ term is excluded). Re-summing this expression to all orders in $j$ reproduces the weak field expansion \eqref{eq:l1b-weak}.
The expansion coefficients $c_n^{(j)}$ can be derived from (\ref{eq:exact2}) and are expressed in terms of gamma functions and Bernoulli numbers \cite{dunne-kogan,Dunne:1999uy}:
\begin{eqnarray}
    c_n^{(j)}=\frac{\Gamma(2n+j+2)\Gamma(2n+j)(2\pi)^{2n+2}\mathcal{B}_{2n+2j+2}}{\Gamma(j+1)\Gamma(2n+3)\Gamma(2n+j+\tfrac{5}{2})}
    \label{eq:cnj}
\end{eqnarray}
Note that these coefficients are factorially divergent as $n\to\infty$ for fixed $j$, and also as $j\to\infty$ for fixed $n$: both the weak field expansion and the gradient expansion are asymptotic series.

The locally constant field approximation analyzed in 
Section \ref{sec:borel} is the leading order of the gradient expansion: the order with no derivatives at all. As we saw, this LCFA expansion is indeed asymptotic, and the large-order growth of the expansion coefficients encodes information about the fluctuations about each instanton sector: recall \eqref{eq:lcfa-b}-\eqref{eq:lcfa}. The same argument extends to an arbitrary order $j$ of the gradient expansion. The leading large order growth of the expansion coefficients is 
\begin{eqnarray}
c_n^{(j)}\sim \qty[(-1)^j\frac{2}{(2\pi)^{2j}\Gamma(j+1)}]\qty(-1)^{n}\Gamma\qty(2n+3j-\tfrac{1}{2})\zeta(2n+2j+2),\qquad n\to \infty,  \quad j\,\, \text{fixed}
\label{eq:cnj-large}
\end{eqnarray}
Given this leading large-order growth, we can sum all orders $j$ of the gradient expansion, and a standard Borel dispersion relation yields the imaginary part of the Borel sum for the corresponding electric field configuration \cite{dunne-kogan}
\begin{eqnarray}
  \Im S(E,\tau)&\sim &L^3 \tau\frac{m^4}{8\pi^{1/2}}\qty(\frac{eE}{\pi m^2})^{5/2}\sum_{k=1}^\infty\frac{1}{k^{5/2}}e^{-k\pi m^2/(eE)}\sum_{j=0}^\infty \frac{1}{j!}\qty(\frac{k\pi m^2}{eE}\frac{\gamma^2}{4})^{j}\\
  &\sim& L^3 \tau\frac{m^4}{8\pi^{1/2}}\qty(\frac{eE}{\pi m^2})^{5/2}\sum_{k=1}^\infty\frac{1}{k^{5/2}}\exp\qty[-\frac{k\pi m^2}{eE}\qty(1-\frac{\gamma^2}{4})]
\end{eqnarray}
The $\gamma$ dependence in the exponent corresponds to the leading small $\gamma$ expansion of the WKB exponent in (\ref{eq:wkb}):
\begin{eqnarray}
  \frac{2}{\sqrt{1+\gamma^2}+1}=1-\frac{\gamma^2}{4}+\ldots
  \label{eq:small-gamma}
\end{eqnarray}

This argument can be extended to include the subleading power-law corrections to the large-order growth in \eqref{eq:cnj-large}:
\begin{equation}
  c_n^{(j)}\sim \qty[(-1)^j\frac{2}{(2\pi)^{2j}\Gamma\qty(j+1)}]\qty(-1)^{n}\Gamma\qty(2n+3j-\tfrac{1}{2})\zeta\qty(2n+2j+2)\sum_{l=0}^\infty\frac{d_l^{(j)}}{\prod_{p=0}^{l-1}\qty(2n+3j-\tfrac{3}{2}-p)}
  \label{eq:cnj2}
\end{equation}
The first three coefficients $d_l^{(j)}$ are listed in table \ref{dcoef}.
\noindent The imaginary part of the analytically continued Borel sum now reads
\begin{align}
  \Im S(E,\tau)&=L^3 \tau\frac{m^4}{8\pi^{1/2}}\qty(\frac{eE}{\pi m^2})^{5/2}\sum_{k=1}^\infty \frac{1}{k^{5/2}}e^{-k\pi m^2/(eE)}\sum_{j=0}^\infty \frac{1}{j!}\qty(\frac{k\pi m^2}{eE}\frac{\gamma^2}{4})^{j} \sum_{n=0}^\infty\qty(\frac{eE}{k\pi m^2})^{\!n} d_n^{(j)}
\end{align}
To simplify this, we convert the polynomials $d_n^{(j)}$ into the form of a sum of falling factorials $(j)_n=j(j-1)(j-2)\ldots(j-n+1)$, listed in the right half of table \ref{dcoef}. Using the identity
\begin{equation}
  \sum_{j=0}^\infty \frac{(j)_n}{j!}x^j=x^n e^x,\quad n\in\mathbbm{Z}_{\geq 0}
\end{equation}
\begin{table}[t]
  \centering
  \setlength{\tabcolsep}{10pt}
  \renewcommand{\arraystretch}{1}
  \caption{The first three coefficients $d_n^{(j)}$ appearing in the large-order growth of the coefficients $c_k^{(j)}$ in \eqref{eq:cnj2}, expressed as a polynomial and as a falling factorial.}
  \label{dcoef}
  \begin{tabular}{c|l|l}
    $n$ & $d_n^{(j)}$ (polynomial) & $d_n^{(j)}$ (falling factorial)\\
    \hline 0 & 1 & 1\\
    1 & $-2 j^2+7 j-\frac{5}{4}$ & $-2 (j)_2+5 (j)_1-\frac{5}{4}(j)_0$\\
    2 & $2 j^4-17 j^3+42 j^2-27 j+\frac{105}{32}$ & $2(j)_4-5(j)_3+5(j)_2+\frac{105}{32}(j)_0$
  \end{tabular}
\end{table}
and denoting the numerical coefficient of the $\ell^\text{th}$-order falling factorial in the $n^\text{th}$ term $d_n^{(j)}$ as $d_{n,\ell}$, we can write
\begin{align}
  \sum_{n=0}^\infty\qty(\frac{eE}{k\pi m^2})^n\sum_{j=0}^\infty \frac{1}{j!}\qty(\frac{k\pi m^2}{eE}\frac{\gamma^2}{4})^{j} d_n^{(j)}&=\sum_{n=0}^\infty\qty(\frac{eE}{k\pi m^2})^n\sum_{j=0}^\infty \frac{1}{j!}\qty(\frac{k\pi m^2}{eE}\frac{\gamma^2}{4})^{j} \sum_{\ell=0}^{2n}d_{n,\ell}(j)_\ell\\
  &=\sum_{n=0}^\infty\sum_{\ell=0}^{2n}d_{n,\ell}\qty(\frac{eE}{k\pi m^2})^n\sum_{j=0}^\infty\frac{(j)_\ell}{j!}\qty(\frac{k\pi m^2}{eE}\frac{\gamma^2}{4})^j\\
  &=\sum_{n=0}^\infty\sum_{\ell=0}^{2n}d_{n,\ell}\qty(\frac{eE}{k\pi m^2})^{n-\ell}\qty(\frac{\gamma^{2}}{4})^{\ell}\exp\qty(\frac{k\pi m^2}{eE}\frac{\gamma^2}{4})
\end{align}
This simplifies the imaginary part to
\begin{equation}
  \Im S(E,\tau)=L^3 \tau\frac{m^4}{8\pi^{1/2}}\qty(\frac{eE}{\pi m^2})^{5/2}\sum_{k=1}^\infty\frac{1}{k^{5/2}}\exp\qty[-\frac{k\pi m^2}{eE}\qty(1-\frac{\gamma^2}{4})]\sum_{n=0}^\infty\sum_{\ell=0}^{2n}d_{n,\ell}\qty(\frac{eE}{k\pi m^2})^{n-\ell}\qty(\frac{\gamma^{2}}{4})^\ell
\end{equation}
To simplify this further, we need knowledge of the exact prefactor and exponent, which follows from the WKB approximation. Introducing these factors into the Borel summation, the fluctuation series reorganizes into
\begin{align}
  \Im S(E,\tau)&=L^3 \tau\frac{m^4}{8\pi^{1/2}}\qty(\frac{eE}{\pi m^2})^{5/2}\qty(1+\gamma^2)^{5/4}\sum_{k=1}^\infty\frac{1}{k^{5/2}} \exp\qty(-\frac{k\pi m^2}{eE}\frac{2}{\sqrt{1+\gamma^2}+1})\sum_{n=0}^\infty\sum_{\ell=0}^\infty \tilde{d}_{n,\ell}\qty(\frac{eE}{k\pi m^2})^n\qty(\frac{\gamma^{2}}{4})^{\ell}\nonumber\\
 &=L^3 \tau\frac{m^4}{8\pi^{1/2}}\qty(\frac{eE}{\pi m^2})^{5/2}\qty(1+\gamma^2)^{5/4}\sum_{k=1}^\infty \frac{1}{k^{5/2}}\exp\qty(-\frac{k\pi m^2}{eE}\frac{2}{\sqrt{1+\gamma^2}+1})\nonumber\\
  &\quad\times\Bigg[1+\qty(\frac{eE}{k\pi m^2})\qty(-\frac{5}{4}+\frac{25 \gamma ^2}{16}-\frac{15 \gamma ^4}{16}+\frac{95 \gamma ^6}{128}-\frac{325 \gamma ^8}{512}+\ldots)\\
  &\quad +\qty(\frac{eE}{k\pi m^2})^2\qty(\frac{105}{32}-\frac{105 \gamma ^2}{64}+\frac{945 \gamma ^4}{512}-\frac{945 \gamma ^6}{512}+\frac{394065 \gamma ^8}{262144}+\ldots)+\ldots\Bigg]\nonumber
\end{align}
The series in $\gamma^2$ which multiply the fluctuation powers of $\left(\frac{e E}{\pi m^2}\right)$ can be recognized as the small $\gamma$ expansions of simple functions proportional to the $b_n^{(1)}(\gamma)$ functions in \eqref{eq:t1exp}:
\begin{equation}
\begin{aligned}
  \Im S(E,\tau)&=L^3 \tau\frac{m^4}{8\pi^{1/2}}\qty(\frac{eE}{\pi m^2})^{5/2}\qty(1+\gamma^2)^{5/4}\sum_{k=1}^\infty \frac{1}{k^{5/2}}\exp\qty(-\frac{k\pi m^2}{eE}\frac{2}{\sqrt{1+\gamma^2}+1})\\
  &\quad\times\Bigg[1-\frac{5}{4}\qty(\frac{eE}{k\pi m^2})\frac{\qty(1-\frac{3}{4}\gamma^2)}{\sqrt{1+\gamma^2}}+\frac{105}{32}\qty(\frac{eE}{k\pi m^2})^2\frac{\qty(1+\frac{1}{4}\gamma^2)^2}{\qty(1+\gamma^2)}+\ldots\Bigg]\label{WKBsubleading}
\end{aligned}
\end{equation}
This  reproduces all $k$-instanton sectors of 
the leading singularity in (\ref{eq:Imexactresurgence}) for all $\gamma$.

\section{Appendix B: Incomplete gamma function representation of the effective action}
\label{sec:appB}

Using Darboux's theorem, we were able to use the large-order growth of the perturbative coefficients in the weak-field expansion for the effective action in \eqref{eq:resurgence-start}. Using the form of the coefficients as they were derived, in particular including the zeta function in the sum, we obtained a Borel representation \eqref{eq:exactresurgence} which was written in terms of the Euler-Heisenberg Borel transform \eqref{eq:he-partial}. This allowed us to easily make contact with the LCFA and WKB approximations, analytically continue to the electric field background \eqref{eq:E}, and understand how the presence of an inhomogeneity modifies the constant-field results. 

The one-loop
effective actions can also be usefully expressed in terms of
incomplete gamma functions using the integral representation
\begin{equation}
    e^x x^{-a}\Gamma\qty(a,x)=\frac{1}{\Gamma(1-a)}\int_0^\infty \dd{s} e^{-s}s^{-a}\frac{1}{x+s}
\end{equation}
For the Euler-Heisenberg effective Lagrangian, in particular the partial fraction expansion \eqref{eq:he-partial} of the Borel transform, this leads to \cite{Jentschura:2001qr}
\begin{equation}
    \mathcal{L}_\text{EH}\qty(\frac{eB}{m^2})=\frac{m^4}{8\pi^2}\qty(\frac{eB}{\pi m^2})^2\sum_{k=1}^\infty \frac{1}{k^2}\qty[\exp\qty(i\frac{k\pi m^2}{eB})\Gamma\qty(0,e^{i\pi/2}\frac{k\pi m^2}{eB})+\exp\qty(-i\frac{k\pi m^2}{eB})\Gamma\qty(0,e^{-i\pi/2}\frac{k\pi m^2}{eB})]
\end{equation}
In this representation, the imaginary part of the effective Lagrangian is given by the discontinuity across the cut of the incomplete gamma functions
\begin{equation}
    e^{-i\pi a}\Gamma\qty(a,e^{i\pi}x)-e^{i\pi a}\Gamma\qty(a,e^{-i\pi}x)=-\frac{2\pi i}{\Gamma(1-a)}
\end{equation}
This requires two separate analytic continuations, $B\mapsto e^{\pm i\pi/2}E$, to reach either side of the branch cuts, with the difference precisely agreeing with the imaginary part of \eqref{eq:l1e}. 

The effective action in the LCFA can also be expressed in terms of incomplete gamma functions, starting from \eqref{LCFAweakfield} and using \eqref{eq:lcfa-b}
\begin{equation}
\begin{aligned}
    S_\text{LCFA}(B,\lambda)&=\lambda L^2T\frac{m^4}{8\pi^{3/2}}\qty(\frac{eB}{\pi m^2})^{5/2}\sum_{k=1}^\infty \frac{1}{k^{5/2}}\sum_{n=0}^\infty \frac{4}{3\pi}(-1)^n\frac{\Gamma\qty(n+\frac{1}{2})\Gamma\qty(n+\frac{5}{2})}{n!}\Gamma\qty(\tfrac{3}{2}-n)\qty(\frac{eB}{k\pi m^2})^n \\
    &\quad\times \Bigg[\qty(e^{i\pi/2})^{3/2-n}\exp\qty(i\frac{k\pi m^2}{eB})\Gamma\qty(n-\frac{1}{2},e^{i\pi/2}\frac{k\pi m^2}{eB})+\\
    &\quad+\qty(e^{-i\pi/2})^{3/2-n}\exp\qty(-i\frac{k\pi m^2}{eB})\Gamma\qty(n-\frac{1}{2},e^{-i\pi/2}\frac{k\pi m^2}{eB})\Bigg]
\end{aligned}
\label{eq:lcfa-gamma}
\end{equation}
We can obtain a similar representation for the full effective action by starting from \eqref{eq:resurgence-start} and expressing the factor $\zeta(2n+4)$ as a sum. The effective action then becomes
\begin{equation}
    S(B,\lambda)\sim \lambda L^2T\,\frac{2m^4}{3\pi^2}\qty(\frac{eB}{\pi m^2})^{\!4}\frac{i}{\qty|s_1|^3}\sum_{k=1}^\infty\frac{1}{k^4}\sum_{\ell=0}^\infty (-i\qty|s_1|)^\ell\frac{b_\ell^{(1)}(\gamma)}{\Gamma\qty(-\tfrac{3}{2}-\ell)}\int_0^\infty \dd{s}e^{-s}s^{1/2-\ell}\sum_{n=0}^\infty (-1)^n\qty(\frac{eBs}{k\pi m^2 \qty|s_1|})^{2n}
\end{equation}
in which the perturbative sum is now simply a geometric series
\begin{equation}
    S(B,\lambda)\sim \lambda L^2T\,\frac{2m^4}{3\pi^2}\qty(\frac{eB}{\pi m^2})^{\!4}\frac{i}{\qty|s_1|^3}\sum_{k=1}^\infty\frac{1}{k^4}\sum_{\ell=0}^\infty (-i\qty|s_1|)^\ell\frac{b_\ell^{(1)}(\gamma)}{\Gamma\qty(-\tfrac{3}{2}-\ell)}\int_0^\infty \dd{s}e^{-s}s^{1/2-\ell}\frac{1}{1+\qty(eBs/(k\pi m^2\qty|s_1|))^2}
\end{equation}
Using the integral representation of the incomplete gamma function given above, we can write the effective action as a sum of incomplete gamma functions (relabeling $\ell\to n$)
\begin{equation}
\begin{aligned}
    S(B,\lambda)&\sim -\lambda L^2T\,\frac{m^4}{3\pi^2}\qty(\frac{eB}{\pi m^2})^{\!5/2}\frac{1}{\qty|s_1|^{3/2}} \sum_{k=1}^\infty\frac{1}{k^{5/2}}\sum_{n=0}^\infty(-i)^n\frac{\Gamma\qty(\frac{3}{2}-n)}{\Gamma\qty(-\tfrac{3}{2}-n)}b_n^{(1)}(\gamma)\qty(\frac{eB}{k\pi m^2})^n\\
    &\quad\times\Bigg[\qty(e^{i\pi/2})^{1/2-n}\exp\qty(i\frac{k\pi m^2\qty|s_1|}{eB})\Gamma\qty(n-\frac{1}{2},e^{i\pi/2}\frac{k\pi m^2\qty|s_1|}{eB})\\
    &\quad -\qty(e^{-i\pi/2})^{1/2-n}\exp\qty(-i\frac{k\pi m^2\qty|s_1|}{eB})\Gamma\qty(n-\frac{1}{2},e^{-i\pi/2}\frac{k\pi m^2\qty|s_1|}{eB})\Bigg]
\end{aligned}
\end{equation}
Normalizing the first coefficient in the sum
\begin{equation}
\begin{aligned}
    S(B,\lambda)&\sim \lambda L^2T\,\frac{m^4}{8\pi^{3/2}}\qty(\frac{eB}{\pi m^2})^{\!5/2}\qty(1+\gamma^2)^{5/4}\sum_{k=1}^\infty\frac{1}{k^{5/2}}\sum_{n=0}^\infty(-i)^n\frac{\Gamma\qty(-\tfrac{3}{2})}{\Gamma\qty(-\tfrac{3}{2}-n)}\frac{b_n^{(1)}(\gamma)}{b_0^{(1)}(\gamma)}\qty(\frac{eB}{k\pi m^2})^n\\
    &\quad\times\Gamma\qty(\tfrac{3}{2}-n)\Bigg[\qty(e^{i\pi/2})^{3/2-n}\exp\qty(i\frac{k\pi m^2\qty|s_1|}{eB})\Gamma\qty(n-\frac{1}{2},e^{i\pi/2}\frac{k\pi m^2\qty|s_1|}{eB})\\
    &\quad +\qty(e^{-i\pi/2})^{3/2-n}\exp\qty(-i\frac{k\pi m^2\qty|s_1|}{eB})\Gamma\qty(n-\frac{1}{2},e^{-i\pi/2}\frac{k\pi m^2\qty|s_1|}{eB})\Bigg]
\end{aligned}
\end{equation}
we see the same fluctuation structure as in (\ref{eq:exactresurgence}). Compared to the Euler-Heisenberg effective Lagrangian, at each order in the (instanton) ``$k$''-sum there is now an infinite sum of incomplete gamma functions, reflecting the appearance of fluctuations about the singularities. The analytic continuation of this representation to the electric field background requires more care compared to the residue calculation discussed earlier, as the imaginary part arises from the discontinuities across the branch cuts of the incomplete gamma functions. In addition, we must rotate $B$ and $\lambda$ in such a way as to keep $\gamma$ real and positive. If we consider the continuations $B\to e^{\pm i\pi/2}E$ and $\lambda\to e^{\mp i\pi/2}\tau$

\begin{equation}
\begin{aligned}
    S\qty(e^{\pm i\pi/2}E,e^{\mp i\pi/2}\tau)&\sim L^3 \tau\,\frac{m^4}{8\pi^{3/2}}\qty(\frac{eE}{\pi m^2})^{\!5/2}\qty(1+\gamma^2)^{5/4}\sum_{k=1}^\infty\frac{1}{k^{5/2}}\sum_{n=0}^\infty (-i)^n\frac{\Gamma\qty(-\tfrac{3}{2})}{\Gamma\qty(-\tfrac{3}{2}-n)}\frac{b_n^{(1)}(\gamma)}{b_0^{(1)}(\gamma)}\qty(\frac{eB}{k\pi m^2})^n\\
    &\quad\times\Gamma\qty(\tfrac{3}{2}-n)\Bigg[\exp\qty(\frac{k\pi m^2\qty|s_1|}{eE})\Gamma\qty(n-\frac{1}{2},\frac{k\pi m^2\qty|s_1|}{eE})\\
    &\quad -e^{\pm i\pi(n-1/2)}\exp\qty(-\frac{k\pi m^2\qty|s_1|}{eE})\Gamma\qty(n-\frac{1}{2},e^{\mp i\pi}\frac{k\pi m^2\qty|s_1|}{eE})\Bigg]
\end{aligned}
\end{equation}
then the difference takes the form
\begin{equation}
    \begin{aligned}
        S\qty(e^{-i\pi/2}E,e^{i\pi/2}\tau)-S\qty(e^{i\pi/2}E,e^{-i\pi/2}\tau)&=-L^3 \tau\,\frac{m^4}{8\pi^{3/2}}\qty(\frac{eE}{\pi m^2})^{\!5/2}\qty(1+\gamma^2)^{5/4}\sum_{k=1}^\infty\frac{1}{k^{5/2}}e^{-k\pi m^2\qty|s_1|/(eE)}\\
        &\quad\times\sum_{n=0}^\infty (-i)^n\frac{\Gamma\qty(-\tfrac{3}{2})}{\Gamma\qty(-\tfrac{3}{2}-n)}\frac{b_n^{(1)}(\gamma)}{b_0^{(1)}(\gamma)}\qty(\frac{eE}{k\pi m^2})^n\Gamma\qty(\tfrac{3}{2}-n)\\
        &\quad\times\Bigg[e^{-i\pi(n-1/2)}\Gamma\qty(n-\frac{1}{2},e^{i\pi}\frac{k\pi m^2\qty|s_1|}{eE})-e^{i\pi(n-1/2)}\Gamma\qty(n-\frac{1}{2},e^{-i\pi}\frac{k\pi m^2\qty|s_1|}{eE})\Bigg]
    \end{aligned}
\end{equation}
Using the discontinuity across the cut of the incomplete gamma function, we immediately obtain the leading imaginary part of the effective action
\begin{equation}
    \begin{aligned}
        \Im S(E,\tau)&:=\frac{1}{2}\Im\qty[S\qty(e^{-i\pi/2}E,e^{i\pi/2}\tau)-S\qty(e^{i\pi/2}E,e^{-i\pi/2}\tau)]\\
        &\sim L^3 \tau\frac{m^4}{8\pi^{1/2}}\qty(\frac{eE}{\pi m^2})^{5/2}\qty(1+\gamma^2)^{5/4}\sum_{k=1}^\infty \frac{1}{k^{5/2}}e^{-k\pi m^2\qty|s_1|/(eE)}\sum_{n=0}^\infty (-i)^n\frac{\Gamma\qty(-\tfrac{3}{2})}{\Gamma\qty(-\frac{3}{2}-n)}\frac{b_n^{(1)}(\gamma)}{b_0^{(1)}(\gamma)}\qty(\frac{eE}{k\pi m^2})^n
    \end{aligned}
\end{equation}
In terms of polylogarithms, we can write
\begin{equation}
    \text{Im} S(E,\tau)\sim L^3\tau \frac{m^4}{8\pi^{1/2}}\qty(\frac{eE}{\pi m^2})^{5/2}\qty(1+\gamma^2)^{5/4}\qty[\text{Li}_{5/2}\qty(e^{-\pi m^2\qty|s_1|/(eE)})-\frac{5}{4}\frac{\qty(1-\frac{3}{4}\gamma^2)}{\sqrt{1+\gamma^2}}\qty(\frac{eE}{\pi m^2})\text{Li}_{7/2}\qty(e^{-\pi m^2\qty|s_1|/(eE)})+\ldots]
\end{equation}
which reduces to the LCFA \eqref{eq:lcfa}
as $\gamma\to 0$. This can now be extended to include the contributions from $s_2$ and $s_3$
\begin{equation}
    \begin{aligned}
        \Im S(E,\tau)&=L^3 \tau\,\frac{m^4}{8\pi^{1/2}}\qty(\frac{eE}{\pi m^2})^{\!5/2}\qty(1+\gamma^2)^{5/4}\sum_{k=1}^\infty\frac{1}{k^{5/2}}\Bigg[e^{-k\pi m^2\qty|s_1|/(eE)}\sum_{n=0}^\infty (-i)^n\frac{\Gamma\qty(-\tfrac{3}{2})}{\Gamma\qty(-\frac{3}{2}-n)}\frac{b_n^{(1)}(\gamma)}{b_0^{(1)}(\gamma)}\qty(\frac{eE}{k\pi m^2})^n\\
        &\quad -2\qty(\frac{\gamma^2}{1+\gamma^2})^{5/4}e^{-k\pi m^2\qty|s_3|/(eE)}\sum_{n=0}^\infty (-i)^n\frac{\Gamma\qty(-\tfrac{3}{2})}{\Gamma\qty(-\frac{3}{2}-n)}\frac{b_n^{(3)}(\gamma)}{b_0^{(3)}(\gamma)}\qty(\frac{eE}{k\pi m^2})^n\\
        &\quad +e^{-k\pi m^2\qty|s_2|/(eE)}\sum_{n=0}^\infty i^n\frac{\Gamma\qty(-\tfrac{3}{2})}{\Gamma\qty(-\frac{3}{2}-n)}\frac{b_n^{(2)}(\gamma)}{b_0^{(2)}(\gamma)}\qty(\frac{eE}{k\pi m^2})^n\Bigg]
    \end{aligned}
\end{equation}
which exactly agrees with the result derived in (\ref{eq:Imexactresurgence}). The corresponding representation in terms of polylogarithms is given in \eqref{eq:Imexactpolylogs}.

\section{Appendix C: Resolving Higher Borel Singularities Using Conformal Maps and Pad\'e}
\label{sec:appC}

\begin{figure}[htb]
  \centering
  \includegraphics[width=\textwidth]{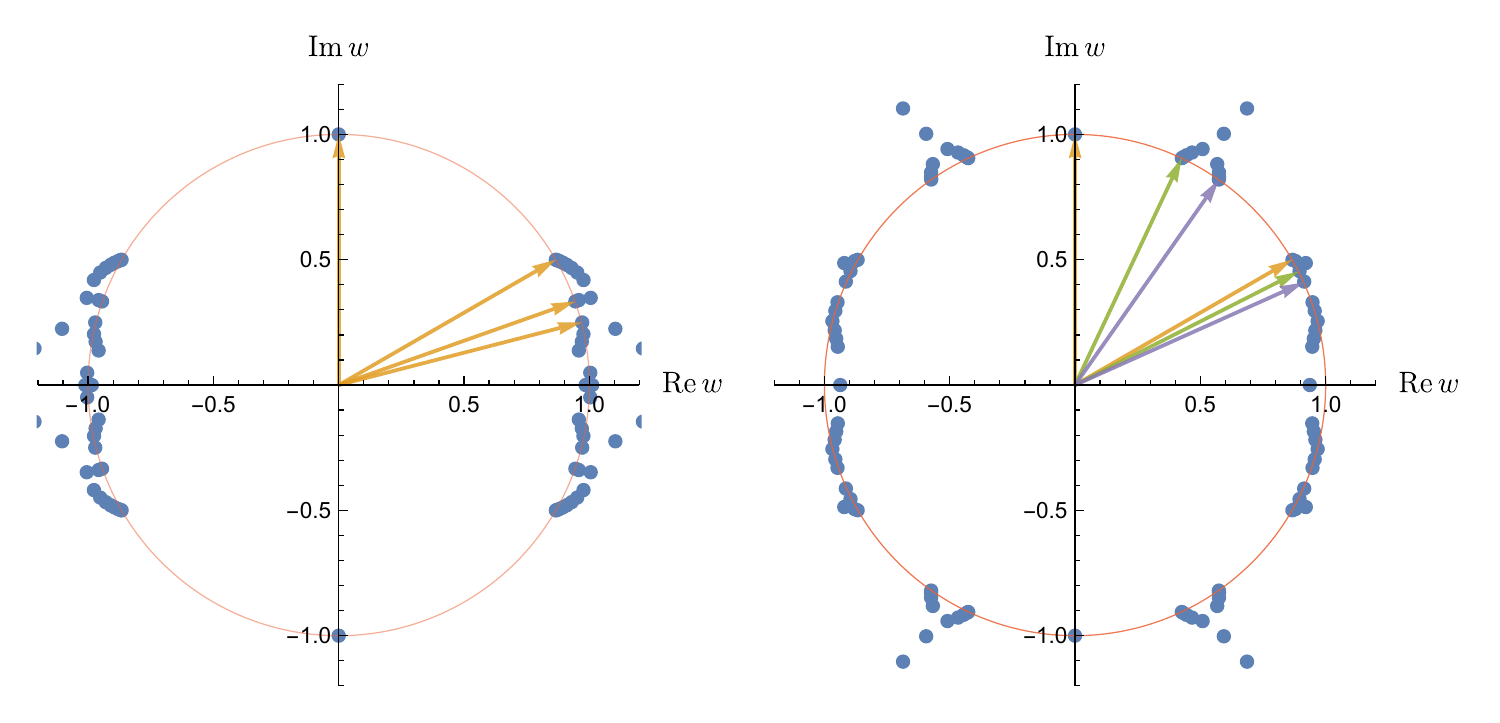}
  \caption{The Pad\'e poles of the conformally-mapped Borel transform in the $w$-plane with $N=100$ and for $\gamma=0.1$ (left) and $\gamma=10$ (right). At small $\gamma$, the first four dominant singularities are the multi-instanton contributions of $\pm s_1$, $\pm 2s_1,\, \pm 3s_1,\, \pm 4s_1$ (orange arrows). Beyond these cannot be cleanly resolved. At large $\gamma$, after $\pm s_1$ (orange arrow) the first sub-leading singularities are $\pm s_3$ (green arrow), followed by $\pm s_2$ (violet arrow). Then we see the contribution of $\pm 2s_1$ (orange arrow), $\pm 2s_3$ (green arrow) and $\pm 2s_2$ (violet arrow).}
  \label{fig:conformal-poles}
\end{figure}

A particular advantage of the Pad\'e-Conformal-Borel method is that it resolves subleading singularities by separating them into the accumulation points of Pad\'e poles on the unit circle in terms of the conformal variable $w$ defined in \eqref{eq:cmap}. The accumulating poles are outside the unit disk, since the function should be analytic inside the disk.\footnote{ Note that the appearance of some poles inside the unit disk occurs for two reasons: first, there is a finite number of terms in our expansion, so that higher singularities (closer to $w=\pm 1$ which represent the point at infinity in the Borel $s$ plane) are resolved imperfectly; and also because the chosen conformal map was only based on the 2 {\it leading} Borel singularities, rather than a full uniformizing map that incorporates the (usually unknown) full singularity structure of the Borel transform function.} This is shown in Figure \ref{fig:conformal-poles}. Using the inverse map, we can identify the location of these singularities in the Borel $s$ plane.  

The left-hand plot in Figure \ref{fig:conformal-poles} is for a small value of the inhomogeneity parameter: $\gamma=0.1$. 
We therefore expect (recall the hierarchy of singularities in \eqref{eq:hierarchy}) that the dominant non-perturbative effects will be due to $s_1$ and its multi-instanton repetitions, rather than the subleading Borel singularities at $s_3$ or $s_2$. Indeed for $\gamma=0.1$ we have the following hierarchy (recall that all the singularities have been normalized by dividing by $|s_1(\gamma)|$):
\begin{eqnarray}
\left[\frac{s_1(\gamma)}{|s_1(\gamma)|}\right]_{\gamma=0.1}
=\pm i \quad &\Rightarrow& \quad w= \pm i
\nonumber\\
\left[\frac{2s_1(\gamma)}{|s_1(\gamma)|}\right]_{\gamma=0.1}
=\pm 2i \quad &\Rightarrow& \quad w= \frac{\sqrt{3}}{2} \pm \frac{1}{2}i
\nonumber\\
\left[\frac{3s_1(\gamma)}{|s_1(\gamma)|}\right]_{\gamma=0.1}
=\pm 3i \quad &\Rightarrow& \quad w= \frac{2\sqrt{2}}{3}\pm  \frac{1}{3}i
\nonumber\\
\left[\frac{4s_1(\gamma)}{|s_1(\gamma)|}\right]_{\gamma=0.1}=\pm 4 i \quad &\Rightarrow& \quad w= \frac{\sqrt{15}}{4} \pm \frac{1}{4} i
\nonumber\\
\vdots &&
\label{eq:hierarchy1}
\end{eqnarray}
The leading singularity at $\pm s_1(\gamma)/|s_1(\gamma)|=\pm i$ maps to $w=\pm i$, and is clearly seen. Since the leading $s_1$ singularity is of square-root type, the conformal map \eqref{eq:cmap} turns it into a pole in terms of $w$, so it appears as an isolated pole in the $w$ plane. 
Due to the hierarchy \eqref{eq:hierarchy1} the first three dominant singularities (those with smallest magnitude in the $s$ plane, or closest on the unit circle to $\pm i$ in the $w$ plane) are $\pm k s_1(\gamma)/|s_1(\gamma)|$ for $k=1,2,3, 4$. These are indicated by the orange arrows. For this small value of $\gamma=0.1$ the $s_3(\gamma)$ and $s_2(\gamma)$ singularities are too distant to be resolved. 

On the other hand, the right-hand plot in Figure \ref{fig:conformal-poles} is for a larger value of the inhomogeneity parameter: $\gamma=10$. Now the dominance hierarchy is very different:
\begin{eqnarray}
\left[\frac{s_1(\gamma)}{|s_1(\gamma)|}\right]_{\gamma=10}
=\pm i \quad &\Rightarrow& \quad w= \pm i
\nonumber\\
\left[\frac{s_3(\gamma)}{|s_1(\gamma)|}\right]_{\gamma=10}
=\pm 1.10499 i \quad &\Rightarrow& \quad w= 0.425438 \pm 0.904988 i
\nonumber\\
\left[\frac{s_2(\gamma)}{|s_1(\gamma)|}\right]_{\gamma=10}=\pm
1.22100 i \quad &\Rightarrow& \quad w= 0.57379 \pm 0.819002 i
\nonumber\\
\left[\frac{2s_1(\gamma)}{|s_1(\gamma)|}\right]_{\gamma=10}
=\pm 2i \quad &\Rightarrow& \quad w= \frac{\sqrt{3}}{2} \pm \frac{1}{2}i
\nonumber\\
\left[\frac{2s_3(\gamma)}{|s_1(\gamma)|}\right]_{\gamma=10}
=\pm 2.20998i \quad &\Rightarrow& \quad w= 0.891768 \pm 0.452494 i
\nonumber\\
\left[\frac{2s_2(\gamma)}{|s_1(\gamma)|}\right]_{\gamma=10}
=\pm 2.44199 i \quad &\Rightarrow& \quad w= 0.91231 \pm 0.409501 i
\nonumber\\
\vdots &&
\label{eq:hierarchy2}
\end{eqnarray}
In this case, the leading singularity at $s_1(\gamma)/|s_1(\gamma)|=\pm i$ is again clearly seen at $w=\pm i$, but the next dominant singularities are $s_3(\gamma)/|s_1(\gamma)|$ and then $s_2(\gamma)/|s_1(\gamma)|$. Only after these two contributions do we see the first multiple of $s_1$. So the next Borel singularities are $2s_1(\gamma)/|s_1(\gamma)|$, $2s_3(\gamma)/|s_1(\gamma)|$ and then $2s_2(\gamma)/|s_1(\gamma)|$. And so on. This confirms that as $\gamma$ increases, the relative dominance between the sub-leading exponential contributions changes, with $s_3(\gamma)$ and $s_2(\gamma)$ becoming more dominant for larger $\gamma$ (i.e., for more inhomogeneous fields). Recall Figure \ref{fig:hierarchy}.


\end{document}